\documentclass[acmsmall]{acmart}
\pdfoutput=1

\acmJournal{PACMPL}
\acmVolume{1}
\acmNumber{POPL} 
\acmArticle{1}
\acmYear{2023}
\acmMonth{1}
\acmDOI{} 
\startPage{1}

\setcopyright{none}

\usepackage{mathtools}
\usepackage{booktabs}
\usepackage{subcaption}
\usepackage[capitalise]{cleveref}
\usepackage{float}
\usepackage{stmaryrd}
\usepackage{xifthen}
\usepackage{tikz}
\usepackage{makecell}

\usetikzlibrary{arrows.meta}

\floatstyle{boxed}
\restylefloat{figure}

\newcommand\mrm[1]{\textrm{#1}}
\newcommand\mbf[1]{\textbf{#1}}
\newcommand\mit[1]{\textit{#1}}

\newcommand\R{\mathbb R}
\newcommand\Z{\mathbb Z}
\newcommand\N{\mathbb N}
\newcommand\ra\rightarrow
\newcommand\lra\multimap
\newcommand\rlra{\mathrel{\mathchoice
  {\overset{\smash{\raisebox{-0.95ex}{\!\footnotesize\textit{R}}}}\rightarrow}
  {\overset{\smash{\raisebox{-0.93ex}{\!\footnotesize\textit{R}}}}\rightarrow}
  {\overset{\smash{\raisebox{-0.7125ex}{\hspace{-1.2pt}\tiny\textit{R}}}}\rightarrow}
  {\overset{\smash{\raisebox{-0.5225ex}{\hspace{-1.2pt}\scalebox{0.7}{\tiny\textit{R}}}}}\rightarrow}%
}}
\newcommand\rlrafootnote{\mathrel{\overset{\smash{\raisebox{-0.95ex}{\!\tiny\textit{R}}}}\rightarrow}}
\newcommand\fun{\lambda}
\newcommand\lfun{\underline\lambda}
\newcommand\rlfun{\underline\lambda_R}
\newcommand\inl{\mrm{inl}}
\newcommand\inr{\mrm{inr}}
\newcommand\case[1]{\mbf{case}\ {#1}\ \mbf{of}}
\newcommand\fst{\mrm{fst}}
\newcommand\snd{\mrm{snd}}
\newcommand\id{\mathsf{id}}
\newcommand\Bool{\ensuremath{\mathsf{Bool}}}
\newcommand\Int{\ensuremath{\mathsf{Int}}}
\newcommand\True{\ensuremath{\mathsf{True}}}
\newcommand\Dual{\mrm{Dual}}
\newcommand\EmptyMap{\mrm{EmptyMap}}
\newcommand\Map{\mrm{Map}}
\newcommand\coloneqqq{\mathbin{\raisebox{0.5pt}{::}{=}}}
\newcommand\Op{\mathsf{Op}}
\newcommand\sem[1]{\llbracket #1 \rrbracket}
\newcommand\trans[2]{\mbf{D}^{#1}_{#2}}
\newcommand\Interleave[1]{\mrm{Interleave}^{#1}}
\newcommand\Deinterleave[1]{\mrm{Deinterleave}^{#1}}
\newcommand\Wrap[1]{\mrm{Wrap}^{#1}}
\newcommand\listappend{\mathbin{+\!\!+}}

\newcommand\Staged{\operatorname*{\mrm{Staged}}}
\newcommand\ZeroStaged{0_{\text{Staged}}}
\newcommand\PlusStaged{\mathbin{+_{\text{Staged}}}}
\newcommand\StagedCall{\mrm{StagedCall}}
\newcommand\InitStaged{\mrm{InitStaged}}
\newcommand\ResolveStaged{\mrm{ResolveStaged}}
\newcommand\StagedMapCot{\mrm{StagedMapCot}}
\newcommand\StagedRunZero{\mrm{StagedRunZero}}

\newcommand\isS{:_S}
\newcommand\isT{:_T}
\newcommand\isB{:_B}
\newcommand\Array{\operatorname*{\mrm{Array}}}
\newcommand\IArray{\operatorname*{\mrm{IArray}}}
\newcommand\StateAlloc{\mrm{StateAlloc}}
\newcommand\InputCot{\mrm{InputCot}}
\newcommand\State{\mrm{State}}
\newcommand\ResolveState{\mrm{ResolveState}}
\newcommand\ur{{!}}
\newcommand\rllet{\underline{\mbf{let}}}
\newcommand\rlin{\underline{\mbf{in}}}
\newcommand\arralloc{\texttt{alloc}}
\newcommand\arrallocBeside{\texttt{allocBeside}}
\newcommand\arrget{\texttt{get}}
\newcommand\arrset{\texttt{set}}
\newcommand\arrsize{\texttt{size}}
\newcommand\arrdealloc{\texttt{dealloc}}
\newcommand\arrmodify{\texttt{modify}}
\newcommand\arrfreeze{\texttt{freeze}}
\newcommand\iarridx{\mathbin{@}}
\newcommand\Contrib{\mrm{Contrib}}
\newcommand\cost{\mrm{cost}}
\newcommand\size{\mrm{size}}

\newcommand\figheading[1]{\textbf{#1}\hfill}

\newcommand\standout[1]{\colorbox{lightgray}{$#1$}}
\newcommand\textstandout[1]{\colorbox{lightgray}{#1}}

\begin{document}

\title{Dual-Numbers Reverse AD, Efficiently}

\author{Tom J.\ Smeding}
\orcid{0000-0002-4986-6820}
\affiliation{%
    \department{Department of Information and Computing Sciences}
    \institution{Utrecht University}
    \city{Utrecht}
    \country{The Netherlands}
}
\email{t.j.smeding@uu.nl}

\author{Matthijs I.\ L.\ V\'ak\'ar}
\orcid{0000-0003-4603-0523}
\affiliation{%
    \department{Department of Information and Computing Sciences}
    \institution{Utrecht University}
    \city{Utrecht}
    \country{The Netherlands}
}
\email{m.i.l.vakar@uu.nl}


\begin{abstract}
Where dual-numbers forward-mode automatic differentiation (AD) pairs each scalar value with its tangent derivative, dual-numbers \emph{reverse-mode} AD attempts to achieve reverse AD using a similarly simple idea: by pairing each scalar value with a backpropagator function.
Its correctness and efficiency on higher-order input languages have been analysed by Brunel, Mazza and Pagani, but this analysis was on a custom operational semantics for which it is unclear whether it can be implemented efficiently.
We take inspiration from their use of \emph{linear factoring} to optimise dual-numbers reverse-mode AD to an algorithm that has the correct complexity and enjoys an efficient implementation in a standard functional language with resource-linear types, such as Haskell.
Aside from the linear factoring ingredient, our optimisation steps consist of well-known ideas from the functional programming community.
Furthermore, we observe a connection with classical imperative taping-based reverse AD, as well as Kmett's \texttt{ad} Haskell library, recently analysed by Krawiec et al.
We demonstrate the practical use of our technique by providing a performant implementation that differentiates most of Haskell98.
\end{abstract}



\maketitle

\section{Introduction}

An increasing number of applications requires computing derivatives of functions specified by a computer program.
The derivative of a function gives more qualitative information of its behaviour around a point (i.e.\ the local shape of the function's graph) than just the function value at that point.
This qualitative information is useful, for example, for optimising parameters along the function (because the derivative tells you how the function changes) or inferring statistics about the function (e.g.\ an approximation of its integral).
These uses appear, respectively, in parameter optimisation in machine learning or numerical equation solving, and in Bayesian inference of probabilistic programs.
Both application areas are highly relevant today.

Automatic differentiation (AD) is the most effective technique for efficient computation of derivatives of programs, and comes in two main flavours: forward AD and reverse AD.
In practice, by far the most common case is that functions have many input parameters and few, or even only one, output parameter; in this situation, forward AD is highly inefficient while reverse AD yields the desired computational complexity.
This is because reverse AD promises to compute the gradient of a function implemented as a program in time at most a constant factor more than runtime of the original program.
However, reverse AD is also significantly more difficult to implement flexibly, correctly and efficiently than forward AD.


Many approaches exist for doing reverse AD on a higher-order language: using taping/tracing in an imperative language (e.g.~\cite{ad-2017-pytorch}) and in a functional language~\cite{ad-2021-kmett-hackage}, using linearisation and transposition code transformations~\cite{dex-2021-ad}, or sometimes specialised by taking advantage of common usage patterns in domain-specific languages~\cite{ad-2022-futhark-partial-recompute}.
In the theory community, various algorithms have been described that apply to a wide variety of source languages, including approaches based on symbolic execution and tracing~\cite{ad-2020-dualnum-revad-linear-factoring,ad-2020-rev-ad-semantics} and on category theory~\cite{vakar-2022-chad}, as well as formalisations of existing implementations~\cite{ad-2021-krawiec-kmett-ad}.
Despite the fact that all these source languages could, theoretically, be translated to a single generic higher-order functional language, each reverse AD algorithm takes a different approach to solve the same problem.
It is unclear how exactly these algorithms relate to each other, meaning that correctness proofs (if any) need to be rewritten for each individual algorithm.

This paper aims to improve on the situation by providing a link from the elegant dual-numbers reverse AD algorithm analysed in~\cite{ad-2020-dualnum-revad-linear-factoring} to a functional taping approach as used in~\cite{ad-2021-kmett-hackage} and analysed in~\cite{ad-2021-krawiec-kmett-ad}.
The key point made by Brunel, Mazza and Pagani~\cite{ad-2020-dualnum-revad-linear-factoring} is that one can attain the right computational complexity by starting from the very elegant dual-numbers reverse AD code transformation (\cref{sec:key-ideas,sec:naive}), and adding a \emph{linear factoring} rule to the operational semantics of the output language of the code transformation.
This linear factoring reduction rule states that for linear functions $f$, the expression $f\ x + f\ y$ should be reduced to $f\ (x + y)$.
Our contributions here are the following:
\begin{itemize}
\item
  We show how the theoretical analysis based on the linear factoring rule can be used as a basis for an algorithm that assumes normal, call-by-value semantics.
  We do this by \emph{staging calls to backpropagators} in \cref{sec:staging}.
\item
  We show how this algorithm can be made efficient by using the standard functional programming techniques of Cayley transformation (\cref{sec:cayley}) and linearly typed functional in-place updates (\cref{sec:mutarrays}).
\item
  Additionally, we explain how the resulting algorithm relates to classical taping-based approaches and the functional taping AD of \cite{ad-2021-kmett-hackage} and \cite{ad-2021-krawiec-kmett-ad} (\cref{sec:improve}).
\item
  Finally, we give an implementation of the final algorithm of \cref{sec:improve-defunctionalisation} that can differentiate most of Haskell98, and that exhibits the correct complexity in practice, as well as reasonable constant-factor performance (\cref{sec:implementation}).
\end{itemize}

\paragraph{Paper overview}
The paper is structured as follows:
\begin{itemize}
\item
  First, in \cref{sec:key-ideas}, we explain the key ideas of this work in a self-contained fashion.
\item
  Afterwards, we start with making precise what we expect of the time complexity of reverse AD in \cref{sec:rev-ad-complexity}, and explore some of the variations on the \emph{type} of a purely-functional code transformation for reverse AD in \cref{sec:rev-ad-type}.
\item
  In \cref{sec:naive} we introduce dual-numbers reverse AD, a code transformation that is simultaneously flexible enough to support e.g.\ full Haskell98\footnote{The changes made in Haskell2010 are mostly irrelevant to differentiation.} (some of the language elements, such as recursive types, are discussed in \cref{sec:source-language-extension}), and also simple enough that it is easy to see intuitively that the output program of the code transformation computes the correct gradient.
  However, there are significant issues with its time complexity; we pinpoint these issues.
\item
  We improve the time complexity of the naive algorithm in \cref{sec:staging} by staging calls to backpropagators (inspired by linear factoring), ensuring that each backpropagator is called at most once.
  However, we note that a problem with the complexity still remains: we are repeatedly adding large structures together in places where we cannot afford performing more than a constant-time operation.
\item
  We solve this problem in \cref{sec:cayley} using the classical ``difference list'' trick, by representing the cotangent accumulator and the monoid of staging maps (over map union) with the Cayley-transformed versions of those monoids, whereby the combination operation becomes function composition: a constant-time operation.
\item
  Furthermore, we observe that these ``monoid updater'' functions are resource-linear, which leads, in \cref{sec:mutarrays}, to a mutable implementation of the cotangent accumulator and staging maps, allowing the removal of the last log-factor in the complexity of the algorithm and a significant reduction in the constant-factor overhead of its runtime.
\item
  After having achieved the correct time complexity, we give some further optimisations in \cref{sec:improve} and link the algorithm to taping and the formalisation of~\cite{ad-2021-krawiec-kmett-ad} in \cref{sec:improve-taping,sec:relation-krawiec}.
\item
  We show how the algorithm naturally generalises to e.g.\ recursion and recursive types in \cref{sec:source-language-extension}.
\item
  Finally, we briefly describe our implementation and show that it has acceptable practical performance in \cref{sec:implementation}, and close off with future work in \cref{sec:future-work} and related work in \cref{sec:related-work}.
\end{itemize}

\subsection*{Acknowledgements}
This project has received funding via NWO Veni grant number VI.Veni.202.124.

\section{Key Ideas}\label{sec:key-ideas}

In traditional dual-numbers \emph{forward} AD, one pairs up the real scalars in the input of a program with their \emph{tangent} (these tangents together form the \emph{directional derivative} of the input), runs the program with overloaded arithmetic operators to propagate forward these tangents, and finally reads the tangent of the output from the tangents paired up with the output scalars of the program.
For example, transforming the program in \cref{subfig:key-example-orig} using dual-numbers forward AD yields \cref{subfig:key-example-fwd}.

\begin{figure}
  \begin{subfigure}[b]{0.24\textwidth}
    \( \begin{array}{l}
        \fun(x : \R, y : \R). \\
        \; \mbf{let}\ z = x + y \\
        \; \mbf{in}\ x \cdot z
    \end{array} \)
    \caption{\label{subfig:key-example-orig}
      The original program
    }
  \end{subfigure}
  \hfill
  \begin{subfigure}[b]{0.32\textwidth}
    \( \begin{array}{l}
        \fun((x : \R, dx : \R) \\
        \hspace{0.23cm},\hspace{-0.03cm} (y : \R, dy : \R)). \\
        \; \mbf{let}\ (z, dz) = (x + y, dx + dy) \\
        \; \mbf{in}\ (x \cdot z, x \cdot dz + z \cdot dx)
    \end{array} \)
    \caption{\label{subfig:key-example-fwd}
      Dual-numbers forward AD
    }
  \end{subfigure}
  \hfill
  \begin{subfigure}[b]{0.42\textwidth}
    \( \begin{array}{l}
        \fun((x : \R, dx : \R \lra (\R, \R)) \\
        \hspace{0.23cm},\hspace{-0.03cm} (y : \R, dy : \R \lra (\R, \R))). \\
        \; \mbf{let}\ (z, dz) = (x + y \\
        \hspace{1.94cm} , \hspace{-0.01cm} \lfun(d : \R).\ dx\ d + dy\ d) \\
        \; \mbf{in}\ (x \cdot z \\
        \hspace{0.54cm} , \hspace{-0.01cm} \lfun(d : \R).\ dz\ (x \cdot d) + dx\ (z \cdot d))
    \end{array} \)
    \caption{\label{subfig:key-example-rev}
      Dual-numbers reverse AD
    }
  \end{subfigure}

  \caption{\label{fig:key-example}
    An example program together with its derivative, both using dual-numbers forward AD and using dual-numbers reverse AD.
    The original program is of type $(\R, \R) \ra \R$.
  }
\end{figure}

For reverse AD, such an elegant formulation is also possible, but we have to somehow encode the ``reversal'' in the tangent scalars that we called $dx$ and $dy$ in \cref{subfig:key-example-fwd}.
A solution is to replace those tangent scalars with \emph{linear functions} that take the \emph{cotangent} (or \emph{reverse derivative}, or \emph{adjoint}) of the scalar it is paired with, and return the cotangent of the full input of the program.\footnote{%
  The actual dualisation process taking place here is taking the \emph{linear negation}: an in some sense more precise type of the program in \cref{subfig:key-example-rev} would be the polymorphic type $\forall c.\ ((\R, \R \lra c), (\R, \R \lra c)) \ra (\R, \R \lra c)$.
  We will use this more precise type in \cref{sec:naive}.
  More discussion can be found in~\cite{ad-2020-dualnum-revad-linear-factoring}.
}
Transforming the same example program \cref{subfig:key-example-orig} using this style of reverse AD yields \cref{subfig:key-example-rev}.
The linearity indicated by the $\lra$-arrow here is that of a vector space\footnote{Although we do not actually use scalar multiplication; see also \cref{sec:rev-ad-type}.} homomorphism; however, operationally, they are just regular functions.

Dual-numbers forward AD has the very useful property that it generalises over many types (e.g.\ products, coproducts, recursive types) and program constructs (e.g.\ recursion, higher-order functions), thereby being applicable to e.g.\ all of Haskell98; the same property is inherited by the style of dual-numbers reverse AD exemplified here.
However, unlike dual-numbers forward AD (which can propagate tangents through a program with only a constant-factor overhead over the original runtime), dual-numbers reverse AD is wildly inefficient; calling $dx_n$ returned by the following differentiated program takes time \emph{exponential} in $n$:
\begin{align*}
  \begin{array}{@{}l@{\ }l@{\ }l@{}}
    \begin{array}{l@{}l}
      \fun&(x_0 : \R, x_1 : \R). \\
        &\mbf{let}\ x_2 = x_0 + x_1 \\
        &\mbf{in}\ \mbf{let}\ x_3 = x_1 + x_2 \\
        &\quad\vdots \\
        &\mbf{in}\ \mbf{let}\ x_n = x_{n-2} + x_{n-1} \\
        &\mbf{in}\ x_n
    \end{array}
    &\rightsquigarrow&
    \begin{array}{l@{}l}
      \fun&((x_0 : \R, dx_0 : \R \ra (\R,\R)), (x_1 : \R, dx_1 : \R \ra (\R,\R))). \\
        &\mbf{let}\ (x_2, dx_2) = (x_0 + x_1, \fun(d : \R).\ dx_0\ d + dx_1\ d) \\
        &\mbf{in}\ \mbf{let}\ (x_3, dx_3) = (x_1 + x_2, \fun(d : \R).\ dx_1\ d + dx_2\ d) \\
        &\quad\vdots \\
        &\mbf{in}\ \mbf{let}\ (x_n, dx_n) = (x_{n-2} + x_{n-1}, \fun(d : \R).\ dx_{n-2}\ d + dx_{n-1}\ d) \\
        &\mbf{in}\ (x_n, dx_n)
    \end{array}
  \end{array}
\end{align*}
Such overhead would make reverse AD completely useless in practice---particularly because other (less flexible) reverse AD algorithms exist that indeed do a lot better.
(See e.g.\ \cite{ad-2018-survey-automatic-differentiation,ad-2018-survey-ad-implementation} for an overview.)

\begin{figure}
  \begin{tikzpicture}
    \node[draw] (1) at (0, 0) {\makecell[c]{
      $\trans1c$ (\cref{fig:algo-naive,fig:wrapper-naive}) \\
      Naive dual-numbers \\ reverse AD
    }};
    \node[draw] (2) at (5.3, 0) {\makecell[c]{
      $\trans2c$ (\cref{fig:staged-tree-impl,fig:algo-staged}) \\
      Stage back- \\ propagator calls \\ with linear \\ factoring${}^\dag$
    }};
    \node[draw] (3) at (10, 0) {\makecell[c]{
      $\trans3c$ (\cref{fig:algo-monadic,fig:staged-tree-impl}) \\
      Use ID generation \\ monad to actually \\ stage backpropagator \\ calls
    }};
    \node[draw] (4) at (1.8, -3.5) {\makecell[c]{
      $\trans4c$ (\cref{fig:algo-cayley,fig:wrapper-cayley}) \\
      Cayley-transformed \\ cotangent collector for \\ cheap zero and plus
    }};
    \node[draw] (5) at (7.8, -3.5) {\makecell[c]{
      $\trans5{}$ (\cref{fig:algo-mutarrays}) \\
      Use mutable arrays with \\ resource-linear types to \\ remove final log-factors \\ from complexity
    }};
    \draw[->] (1) -- (2)
      node [midway, above] {\footnotesize \vphantom{a}\smash{Replace top-level}}
      node [midway, below] {\footnotesize \vphantom{a}\smash{cotangent type}};
    \draw[->] (2) -- (3)
      node [midway, above] {\footnotesize \vphantom{a}\smash{To monadic}}
      node [midway, below] {\footnotesize \vphantom{a}\smash{code}};
    \draw[->] (3) .. controls (9, -2.7) and (3, -1.2) .. (4)
      node [midway, above] {\footnotesize Cayley-transform};
    \draw[->] (4) -- (5)
      node [midway, above] {\footnotesize \vphantom{a}\smash{Resource-linear}}
      node [midway, below] {\footnotesize \vphantom{a}\smash{types}};
  \end{tikzpicture}

  \caption{\label{fig:key-overview}
    Overview of the optimimsations to dual-numbers reverse AD as a code transformation that are described in this paper.
    ($\dag$ = inspired by~\cite{ad-2020-dualnum-revad-linear-factoring})
  }
\end{figure}
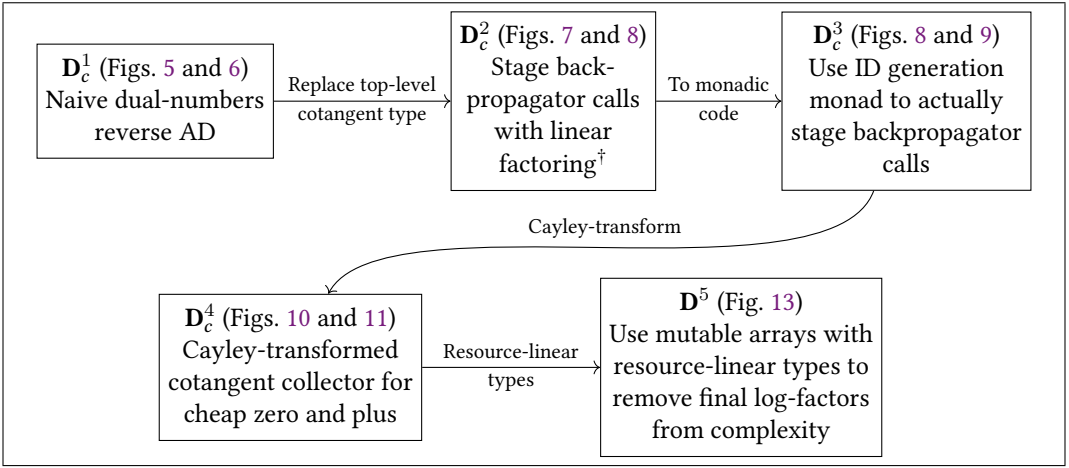

However, it turns out that this form of dual-numbers reverse AD can be \emph{optimised} to be as efficient (in terms of time complexity) as these other algorithms---and most of these optimisations are just applications of standard functional programming techniques.
This paper presents a sequence of changes to the code transformation (see the schematic overview in \cref{fig:key-overview}) that fix all the complexity issues and, in the end, produce an algorithm with which the differentiated program has only a constant-factor overhead in runtime over the original program.
This complexity is as desired from a reverse AD algorithm, and is best possible, while nevertheless being applicable to a wide range of programming language features.

The added value of our deduction over existing publications about reverse AD algorithms is that we show how, by applying a sequence of fairly standard optimisation steps, the functional dual-numbers reverse AD algorithm of~\cite{ad-2020-dualnum-revad-linear-factoring,ad-2021-dual-revad-linear-factoring-pcf,ad-2020-sam-mathieu-matthijs} effectively reduces to a classical taping approach that that is known to have good practical efficiency~\cite{ad-2017-pytorch,ad-2019-tensorflow-eager}.

\paragraph{The optimisation steps}
Let us first intuitively present the optimisation steps as outlined in \cref{fig:key-overview}, before diving into the details.
The initial dual-numbers reverse AD transformation in \cref{fig:algo-naive} is simple and it is easy to see why it is correct via a logical relations argument~\cite{nunes-2022-dual-numbers}.
The idea of this argument is to prove via induction that a backpropagator $x' : \R \lra c$ that is paired with an intermediate value $x : \R$ in the program, holds the gradient of the computation that calculates $x : \R$ from the global input of type $c$.

As explained above, we need to fix the complexity of this algorithm, and the first thing we do is applying \emph{linear factoring}: for a linear function $f$, such as a backpropagator, by definition we have that $f\ x + f\ y = f\ (x + y)$, else it would not be linear.
Observing the form of all the backpropagators in \cref{fig:algo-naive}, we see that in the end all we produce is a giant sum of applications of backpropagators to scalar values; hence, in this giant sum, we should be able to contract applications to the same backpropagator using this linear factoring rule.

We do this linear factoring by not returning a plain $c$ (presumably the type of the program input) from our backpropagators, but instead a $c$ wrapped in an object that can delay calls to linear functions producing a $c$.
This object we call $\Staged$; in the updated code transformation in \cref{fig:algo-staged}, we see that aside from changing the monoid that we are mapping into from $(c, \underline0, (+))$ to $(\Staged c, \ZeroStaged, (\PlusStaged))$, the only material change is that the calls to $d_i$ in $\trans2c[\mit{op}]$ are now wrapped using $\StagedCall$, which delays the calls to $d_i$ by storing the relevant metadata in the returned $\Staged$ object.

However, we notice that we cannot actually implement $\trans2c$ because the required interface of this $\Staged$ object cannot be implemented: it does not make sense to use (linear) function values as keys in a tree map, which the $\Staged$ interface for $\trans2c$ prescribes.
We solve this problem by generating a unique identifier (ID) for each backpropagator that we create, which we do by changing the code transformation to work monadically in an ID generation monad (a special case of a state monad).
The result is shown in \cref{fig:algo-monadic}, which is identical to the previous version in \cref{fig:algo-staged} apart from threading through the next-ID-to-generate.
(The code looks very different, but this is only due to monadic bookkeeping.)
This version, $\trans3c$, \emph{can} be implemented.

At this point, the code transformation reaches a significant milestone: by staging (delaying) calls to backpropagators as long as possible, we have achieved that \emph{every backpropagator is called at most once} (and exactly once if the originating source program fragment was not dead code in the first place).
This milestone is achieved using the following observation: if we assign incrementing numeric IDs at runtime to lambda functions in a functional program, then the runtime closure of a lambda function can only refer to other functions with \emph{smaller} IDs.
Since furthermore our backpropagators only call other functions contained in their closure (and not functions contained in their input argument), we know that all backpropagator calls are to other backpropagators with smaller IDs!
This allows us to resolve all backpropagator calls in the $\Staged$ object produced by the differentated program by simply calling them from highest ID to lowest ID, and collecting and combining (using linear factoring) the calls made along the way.
For more details on this observation and the resulting resolve algorithm, we refer to \cref{sec:staged-interface}.

But we are not done yet.
The code transformation at this point ($\trans3c$ in \cref{fig:algo-monadic}) still has a glaring problem: orthogonal to the issue that backpropagators were called too many times (which we fixed), we are still creating one-hot input cotangent values and adding those together.
This problem is somewhat more subtle, because it is not actually apparent in the program transformation itself; indeed, looking back at \cref{subfig:key-example-rev}, there are no one-hot values to be found.
However, the only way to \emph{use} the program in \cref{subfig:key-example-rev} to do something useful, namely to compute the cotangent (gradient) of the input, is to pass $(\lambda z.\ (z, 0))$ to $\mit{dx}$ and $(\lambda z.\ (0, z))$ to $\mit{dy}$; it is easy to see that generalising this to larger input structures, which we do in the full development below, results in input values like $(0, \ldots, 0, z, 0, \ldots, 0)$ that get added together.
Adding many zeros together can hardly be the most efficient way to go about things, and indeed this is a complexity issue in the algorithm.

The way we solve this problem of one-hots is less AD-specific: the two optimisations that we perform are Cayley-transformation and exploiting resource-linearity.
Cayley-transformation is a classic technique in functional programming also known as \emph{difference lists}~\cite{fp-1986-difference-lists} in the Haskell community.
Resource-linearity draws from classic theory about linear logic but has only relatively recently been popularised among the general programmer community, and can now be used in fairly well-known languages like Rust\footnote{\url{https://www.rust-lang.org}} and Haskell~\cite{fp-2018-linear-haskell}.
While resource-linearity is not essential in our work (we could just as well have used a monad such as the \texttt{ST} monad in Haskell~\cite{fp-1994-st-monad}), we feel that this results in a nicer presentation that more faithfully depicts which operations require strict sequentiality and which do not.
The Cayley-transformation brings the code transformation in precisely the right shape to make good use of resource-linearity (in \cref{fig:array-interface,fig:algo-mutarrays}).

Having fixed the second problem (of one-hots), we have now obtained a code transformation with the right complexity: the differentiated program can compute the gradient of the source program at some input with runtime proportional to the runtime of the source program.
We are now also in the position to note the similarity to taping-based AD as in \cite{ad-2021-kmett-hackage,ad-2021-krawiec-kmett-ad}: the incrementing IDs that we attached to backpropagators earlier essentially give a mapping from $\{0,\ldots,n\}$ to our backpropagators.
Furthermore, each backpropagator corresponds to either a primitive arithmetic operation performed in the source program, or to an input value; this already means we have a tape, in a sense, of all performed primitive operations, albeit in the form of a chain of closures.
The optimisation using resource-linearity then eliminates also this last difference, because there we actually reify this tape in a large array.

\section{Preliminaries: The complexity of reverse AD}\label{sec:rev-ad-complexity}

The only reason, in practice, for using reverse AD over forward AD (which is significantly easier to implement) is computational complexity.
Arguably, therefore, it is important that we fix precisely what the time complexity of reverse AD ought to be, and check that any proposed algorithm indeed conforms to this time complexity.

In this paper we discuss a code transformation, so we phrase the desired time complexity in terms of a code transformation $\mathcal R$ that takes a program $P$ of type $\R^n \ra \R^m$ to a program $\mathcal R[P]$ of type\footnote{We will discuss the type of reverse AD in more detail in \cref{sec:rev-ad-type}, generalised beyond vectors of reals.} $(\R^n, \underline{\R^m}) \ra \underline{\R^n}$ that computes the reverse derivative of $P$.
The classic result (see ~\cite{adbook-2008-griewank-walther}) is that first first-order languages, $\mathcal R$ exists such that the following criterion is satisfied:
\begin{align*}
  &\exists c > 0.\ \forall P : \mrm{Programs}(\R^n \ra \R^m).\ \forall I : \R^n, A : \underline{\R^m}. \\
  &\qquad \cost(\mathcal R[P]\ (I, A))
    \leq c \cdot (\cost(P\ I) + \size(I))
\end{align*}
where we denote by $\cost(E)$ the amount of time required to evaluate the expression $E$ to normal form, and by $\size(I)$ the amount of time required to read all of $I$ sequentially.
(Note that $\cost(\mathcal R[P]\ (I, A))$ does not measure the cost of evaluating the code transformation $\mathcal R$ itself; that is considered to be a compile-time cost.)
In particular, if $P$ reads its entire input (and does not ignore part of it), the second line can be simplified to $\cost(\mathcal R[P]\ (I, A))) \leq c \cdot \cost(P\ I)$.

The most important point of this criterion is that $c$ cannot depend on $P$: informally, the output program produced by reverse AD is not allowed to have more than a constant factor overhead in runtime over the original program, and this constant factor is uniform over all programs.

A weaker form of the criterion is sometimes used (as e.g.\ in~\cite{ad-2022-futhark-partial-recompute}, where $c$ is proportional to the largest scope depth in the program) where $c$ is dependent on the program in question but not on the size of the input to that program; this can only make sense in a language that has variably sized arrays or similar structures.
In this case, the criterion is expressed for a program family $\mit{PF}$ that should be understood to be the same program for all $n$, just with different input sizes:
\begin{align*}
  &\forall \mit{PF} : (n : \N) \ra \mrm{Programs}(\R^n \ra \R^m).\ \exists c > 0.\ \forall n \in \N.\ \forall I : \R^n, A : \underline{\R^m}. \\
  &\qquad \mrm{cost}(\mathcal R[\mit{PF}_n]\ (I, A)) \leq c \cdot (\cost(\mit{PF}_n\ I) + \size(I))
\end{align*}
where $c$ is preferably at most linearly or sub-linearly dependent on the size of the program code of $\mit{PF}$.

The final version of the code transformation described in this paper satisfies the first (most stringent) criterion.

\section{Preliminaries: The type of reverse AD}\label{sec:rev-ad-type}

Before one can define an algorithm, one has to fix the type of that algorithm.
For forward AD on first-order programs (or at least, programs for which the input and output does not contain function values), the desired type seems quite evident: $\mathcal F : (a \ra b) \rightsquigarrow ((a, \underline{a}) \ra (b, \underline{b}))$, where we write $T_1 \rightsquigarrow T_2$ for a (compiler) code transformation taking a program of type $T_1$ and returning a program of type $T_2$, and where $\underline{a}$ is the type of tangent vectors (derivatives) of values of type $a$.
This distinction between the type of values $a$ and the type of their derivatives $\underline{a}$ is important in some versions of AD, but will be mostly cosmetic in this paper; in an implementation one can take $\underline{a} = a$, but there is some freedom in this choice.\footnote{In particular, $\underline{\R} = \R$, but for $\underline{\texttt{Int}}$ one can choose the unit type $()$ and be perfectly sound and consistent.}
Given a program $f : a \ra b$, $\mathcal F[f]$ is a program that takes, in addition to its regular argument, also that argument's derivative; the output is then the regular result paired up with the derivative of that result.

More specifically, for forward AD, we want the following in the case that $a = \R^n$ and $b = \R^m$ (writing $\mbf{x} = (x_1, \ldots, x_n)$):\footnote{This generalises to more complex (but still first-order) in/outputs by regarding those as collections of real values as well.}
\begin{align*}
\mathcal F[f]\left(\mbf{x}, \left(\frac{\partial x_1}{\partial\alpha}, \ldots, \frac{\partial x_n}{\partial \alpha}\right)\right)
= \left(f(\mbf{x}), \left(\frac{\partial f(\mbf{x})_1}{\partial\alpha}, \ldots, \frac{\partial f(\mbf{x})_m}{\partial\alpha}\right)\right)
\end{align*}
Setting $\alpha = x_i$ means passing $(0, \ldots, 1, \ldots, 0)$ as the argument of type $\underline{a}$ and computing the partial derivative with respect to $x_i$ of $f(\mbf{x})$.
In other words, $\snd(\mathcal F[f](x, \mit{dx}))$ is the directional derivative of $f$ at $x$ in the direction $\mit{dx}$.

For reverse AD, the desired type is less evident.
A first guess would be:
\begin{align*}
\mathcal R_1 : (a \ra b) \rightsquigarrow ((a, \underline{b}) \ra (b, \underline{a}))
\end{align*}
with the following intended meaning for $a = \R^n$ and $b = \R^m$ (again writing $\mbf{x} = (x_1, \ldots, x_n)$):
\begin{align*}
\mathcal R_1[f]\left(\mbf{x}, \left(\frac{\partial\omega}{\partial f(\mbf{x})_1}, \ldots, \frac{\partial\omega}{\partial f(\mbf{x})_m}\right)\right)
= \left(f(\mbf{x}), \left(\frac{\partial\omega}{\partial x_1}, \ldots, \frac{\partial\omega}{\partial x_n}\right)\right)
\end{align*}
In particular, if $b = \R$ and we pass 1 as its reverse derivative (cotangent, adjoint) of type $\underline{b}$, we obtain the gradient of the input in the $\underline{a}$-typed output.

However, $\mathcal R_1$ is not readily implementable for even moderately interesting languages.
One way to see this is to acknowledge the reality that the type $\underline{a}$ (of derivatives of values of type $a$) should really be dependent on the input value, or \emph{primal value}, of type $a$.
Let us write it instead as $\mathcal D[a](x)$, where $x : a$ is that primal value.
For scalars and product types this dependence does not yet occur (i.e.\ $\mathcal D[a](-)$ is independent of its argument), but for sum types (coproducts) it does: the only reasonable derivatives for a value $\inl(x) : \sigma + \tau$ (for $x : \sigma$) are of type $\underline{\sigma}$.
Letting $\underline{\sigma + \tau} = \underline{\sigma} + \underline{\tau}$ would allow passing a derivative value of type $\underline{\tau}$ to $\inl(x) : \sigma + \tau$, which is nonsensical (and an implementation could do little else than throw a runtime error in that situation).
The derivative of $\inl(2x) : \R + \Bool$ cannot be $\inr(\True)$; it should at least somehow contain a real value.

Similarly, the derivative for a dynamically sized array, if the input language supports those, must really be of the same size as the input array.
This, too, is a dependence of the type of the derivative on the \emph{value} of the input.

Therefore, the output type of forward AD which we wrote above as $(a, \underline{a}) \ra (b, \underline{b})$ should really be $(\Sigma_{x : a}\, \mathcal D[a](x)) \ra (\Sigma_{y : b}\, \mathcal D[b](y))$, rendering what were originally pairs of value and tangent now as \emph{dependent} pairs of value and tangent.
This is a perfectly sensible type, and indeed correct for forward AD, but it does not translate at all well to reverse AD in the form of $\mathcal R_1$: the output type would be something like $(\Sigma_{x : a}\, \mathcal D[b](y)) \ra (\Sigma_{y : b}\, \mathcal D[a](x))$, which is nonsense because both $x$ and $y$ are out of scope.

Another, less mathematically heavy way to see that the type of $\mathcal R_1$ is unusable, is to note that one cannot even differentiate let-bindings using $\mathcal R_1$.
In order to apply to (an extension of) the lambda calculus, let us rewrite the types somewhat, representing what we previously wrote as a function $f : a \ra b$ as a term $x : a \vdash t : b$ having its input in a free variable and producing its output as the returned value.
Making the modest generalisation to support any full environment as input (instead of just a single variable), we get $\mathcal R_1 : (\Gamma \vdash t : a) \rightsquigarrow (\Gamma, d : \underline{a} \vdash \mathcal R_1[t] : (a, \underline{\Gamma}))$, where $\underline{\Gamma}$ is a tuple containing the derivatives of all elements in the environment $\Gamma$.
We define $\underline{\varepsilon} = ()$ for the empty environment and $\underline{\Gamma, x : a} = (\underline{\Gamma}, \underline{a})$ inductively.

Now, consider differentiating the following program using $\mathcal R_1$:
\begin{align*}
\Gamma \vdash (\mbf{let}\ x = e_1\ \mbf{in}\ e_2) : a
\end{align*}
where $\Gamma \vdash e_1 : b$ and $\Gamma, x : b \vdash e_2 : a$.
Thus, $\mathcal R_1$ needs to somehow build a program of this type:
\begin{align}
\Gamma, d : \underline{a} \vdash \mathcal R_1[\mbf{let}\ x = e_1\ \mbf{in}\ e_2] : (a, \underline{\Gamma})
\label{r1_deriv_letbind}
\end{align}
However, recursively applying $\mathcal R_1$ on $e_1$ and $e_2$ yields terms:
\begin{align*}
\Gamma, d : \underline{b} &\vdash \mathcal R_1[e_1] : (b, \underline{\Gamma}) \\
\Gamma, x : b, d : \underline{a} &\vdash \mathcal R_1[e_2] : (a, (\underline{\Gamma}, \underline{b}))
\end{align*}
To produce the program in \eqref{r1_deriv_letbind}, we cannot use $\mathcal R_1[e_2]$ because we do not yet have an $x : b$ (which needs to come from $\mathcal R_1[e_1]$), and we cannot use $\mathcal R_1[e_1]$ because the $\underline{b}$ needs to come from $\mathcal R_1[e_2]$!
The type of $\mathcal R_1$ demands the cotangent of the result \emph{too early}.

Of course, one might argue that we can just use $e_1$ to compute the $b$, $\mathcal R_1[e_2]$ to get the $\underline{b}$ and $e_2$'s contribution to $\underline{\Gamma}$, and finally $\mathcal R_1[e_1]$ to get $e_1$'s contribution to $\underline{\Gamma}$ based on its own cotangent of type $\underline{b}$.
However, this would essentially compute $e_1$ twice (once directly and once as part of $\mathcal R_1[e_1]$), meaning that the time complexity becomes super-linear in the depth of let-bindings, which is unacceptable.

So apart from not being precisely typeable, $\mathcal R_1$ is also not implementable in a compositional way.

\paragraph{Fixing the type of reverse AD}
Both when looking at the dependent type of $\mathcal R_1$ and when looking at its implementation, we found that the cotangent $\mit{dy} : \mathcal D[b](y)$ was required before the result $y : b$ was itself computed.
One way to solve this issue is to just postpone requiring the cotangent of $y$, i.e.\ to instead look at $\mathcal R_2$:
\begin{align*}
\begin{array}{l@{\qquad}c}
  \mathcal R_2 : (a \ra b) \rightsquigarrow (a \ra (b, \underline{b} \ra \underline{a}))
    & \text{(non-dependent version)} \\
  \mathcal R_2 : (a \ra b) \rightsquigarrow (\Pi_{x : a}\, \Sigma_{y : b}\, (\mathcal D[b](y) \ra \mathcal D[a](x)))
    & \text{(dependent version)}
\end{array}
\end{align*}
Note that this type \emph{is} well-scoped.
Furthermore, this ``derivative function'' mapping the cotangent of the result to the cotangent of the argument is actually a linear function, in the sense of a vector space homomorphism: indeed, it is multiplication by the Jacobian matrix of $f$, the input function.
This way, we get:
\begin{align*}
\mathcal R_2 : (a \ra b) \rightsquigarrow (\Pi_{x : a}\, \Sigma_{y : b}\, (\mathcal D[b](y) \lra \mathcal D[a](x)))
\end{align*}
which is the type of the reverse AD code transformation derived by \cite{adfp-2018-categories-ad} and in CHAD (\cite{vakar-2021-higher-order-reverse-ad}, extended and improved in \cite{vakar-2022-chad}, generalised in \cite{nunes-2022-chad-expressive}).\footnote{Actually, CHAD has non-identity type mappings for the primal types $x : a$ and $y : b$ as well in order to compositionally support function values in a way that fits the type of $\mathcal R_2$. We consider only the top-level type in this discussion, and for first-order in- and output types, the two coincide.}

While this formulation of reverse AD admits a rich mathematical foundation and a straightforward program transformation on the lambda calculus, it is tricky to make efficient.

\paragraph{Applying Yoneda}
We can further transform $\mathcal R_2$ to obtain a type for reverse AD that is somewhat reminiscent of dual-numbers forward AD in $\mathcal F$.
Applying the Yoneda lemma on the $\lra$-arrow in the type of $\mathcal R_2$, we get $\mathcal R_3'''$ below; weakening this type somewhat by enlarging the scope of the $\forall c$ quantifier, weakening some more by moving back the $(\mathcal D[a](x) \lra c)$ argument, and finally uncurrying, we arrive at $\mathcal R_3$:
\begin{align*}
\mathcal R_3''' &: (a \ra b) \rightsquigarrow \Pi_{x : a}\, \Sigma_{y : b}\, \forall c.\, ((\mathcal D[a](x) \lra c) \ra (\mathcal D[b](y) \lra c)) \\
\mathcal R_3'' &: (a \ra b) \rightsquigarrow \forall c.\, \Pi_{x : a}\, \Sigma_{y : b}\, ((\mathcal D[a](x) \lra c) \ra (\mathcal D[b](y) \lra c)) \\
\mathcal R_3' &: (a \ra b) \rightsquigarrow \forall c.\, \Pi_{x : a}\, ((\mathcal D[a](x) \lra c) \ra \Sigma_{y : b}\, (\mathcal D[b](y) \lra c)) \\
\mathcal R_3 &: (a \ra b) \rightsquigarrow \forall c.\, ((\Sigma_{x : a}\, (\mathcal D[a](x) \lra c)) \ra \Sigma_{y : b}\, (\mathcal D[b](y) \lra c)) \\
&\hphantom{: (a \ra b) \;\rightsquigarrow\;} \color{gray} \forall c.\, ((a, \hspace{1.5em} \underline{a} \hspace{3em} \lra c\hspace{0.35em}) \ra (b, \hspace{1.07em} \underline{b} \hspace{3em} \lra c))
\end{align*}

The $\lra$-arrows in these types are really vector space homomorphisms, but since we will only use the substructure of commutative monoids in this paper (and forget about scalar multiplication), we will always consider $\lra$-functions commutative monoid homomorphisms.

In any case, the $\lra$-arrows in these types, as well as the $c$ bound by the $\forall$-quantifier, live in the category of commutative monoids.
Indeed, $c$ will always have a commutative monoid structure in this paper; that is: it has a zero as well as a commutative, associative addition operation $(+) : (c, c) \ra c$.

Returning to the types in question, we see that we can convert $\mathcal R_2[t]$ to $\mathcal R_3[t]$:
\begin{align*}
&(\fun (x : a, \mit{dx} : \mathcal D[a](x) \lra c). \\
&\hspace{3em} \mbf{let}\ (y : b, \mit{dy} : \mathcal D[b](y) \lra \mathcal D[a](x)) = \mathcal R_2[t]\ x\ \mbf{in}\ (y, \mit{dx} \circ \mit{dy})) \\
&\hspace{1em} : \forall c.\, ((\Sigma_{x : a}\, (\mathcal D[a](x) \lra c)) \ra \Sigma_{y : b}\, (\mathcal D[b](y) \lra c))
\end{align*}
And we can also convert $\mathcal R_3[t]$ back to $\mathcal R_2[t]$, but due to how we weakened the types above, only in the non-dependent world:
\begin{align*}
&(\fun (x : a).\ \mathcal R_3[t]\ (x, \id))
    : a \ra (b, \underline b \lra \underline a)
\end{align*}
So in some sense, $\mathcal R_2$ and $\mathcal R_3$ compute the same thing, albeit with types that differ somewhat in their accurate portrayal of the type dependencies.

In fact, $\mathcal R_3$ admits a very elegant implementation as a program transformation that is structurally recursive over all language elements except for the primitive operations in the leaves.
However, there are some issues with the computational complexity of this straightforward implementation of $\mathcal R_3$, one of which we will fix here immediately, and the other of which is the topic of the rest of this paper.

\paragraph{Moving the pair to the leaves}
Let us return to forward AD for a moment.
Recall the type we gave for forward AD:\footnote{We revert to the non-dependent version for now because the dependencies are irrelevant for this point, and they clutter the presentation.}
\begin{align*}
\mathcal F : (a \ra b) \rightsquigarrow ((a, \underline{a}) \ra (b, \underline{b}))
\end{align*}
Supposing we have a program $f : ((\R_1, \R_2), \R_3) \ra \R_4$, we get: (the subscripts are semantically meaningless and are just for tracking arguments)
\begin{align*}
\mathcal F[f] : (((\R_1, \R_2), \R_3), ((\underline{\R_1}, \underline{\R_2}), \underline{\R_3})) \ra (\R_4, \underline{\R_4})
\end{align*}
While this is perfectly implementable and correct and efficient, it is not the type that corresponds to what is by far the most popular implementation of forward AD, namely \emph{dual-numbers forward AD}, which has the following type:
\begin{gather*}
\mathcal F_{\text{dual}} : (a \ra b) \rightsquigarrow (\Dual[a] \ra \Dual[b]) \\
\Dual[\R] = (\R, \underline\R) \qquad
\Dual[()] = () \qquad
\Dual[(a, b)] = (\Dual[a], \Dual[b])
\end{gather*}
Intuitively, instead of putting the pair at the root like $\mathcal F$ does, $\mathcal F_{\text{dual}}$ puts the pair at the leaves---more specifically, at the scalars in the leaves, leaving non-$\R$ types like $()$ or $\Int$ alone.
For the given example program $f$, dual-numbers forward AD would yield the following the derivative program type:
\begin{align*}
\mathcal F_{\text{dual}}[f] : (((\R_1, \underline{\R_1}), (\R_2, \underline{\R_2})), (\R_3, \underline{\R_3})) \ra (\R_4, \underline{\R_4})
\end{align*}
Of course, for any given types $a, b$ the two versions are trivially inter-converted, and as stated, for forward AD both versions can be defined inductively equally well, resulting in efficient programs in terms of time complexity.

However, for reverse AD in the style of $\mathcal R_3$, the difference between $\mathcal R_3$ and its pair-at-the-leaves dual-numbers variant ($\mathcal R_{3\text{dual}}$ below) is more pronounced.
First note that indeed both styles (with the pair at the root and with the pair at the leaves) produce a sensible type for reverse AD: (again for $f : ((\R_1, \R_2), \R_3) \ra \R_4$)
\begin{align*}
\mathcal R_3[f] &: \forall c.\, (((\R_1, \R_2), \R_3), ((\underline{\R_1}, \underline{\R_2}), \underline{\R_3}) \lra c) \ra (\R_4, \underline{\R_4} \lra c) \\
\mathcal R_{3\text{dual}}[f] &: \forall c.\, (((\R_1, \underline{\R_1} \lra c), (\R_2, \underline{\R_2} \lra c)), (\R_3, \underline{\R_3} \lra c)) \ra (\R_4, \underline{\R_4} \lra c)
\end{align*}
The individual functions of type $\underline{\R} \lra c$ are usually called \emph{backpropagators} in literature, and we will adopt this terminology.

Indeed, these two programs are again easily inter-convertible, if one realises that:
\begin{enumerate}
\item $c$ is a commutative monoid and thus posesses an addition operation, which can be used to combine the three $c$ results into one for producing the input of $\mathcal R_3$ from the input of $\mathcal R_{3\text{dual}}$;
\item The function $g : ((\underline{\R_1}, \underline{\R_2}), \underline{\R_3}) \lra c$ is linear, and hence e.g.\ $\fun (x : \underline{\R_2}).\ g\ ((0, x), 0)$ suffices as value for $\underline{\R_2} \lra c$.
\end{enumerate}

However, the problem arises when defining $\mathcal R_3$ inductively as a program transformation.
To observe this difference between $\mathcal R_3$ and $\mathcal R_{3\text{dual}}$, consider the term $t = \fun(x : (a, b)).\ \fst(x)$ of type $(a, b) \ra a$ and the types of its derivative using both methods:
\begin{gather*}
\begin{array}{r@{\ }l}
\mathcal R_3[t] &: \forall c.\ ((a, b), (\underline{a}, \underline{b}) \lra c) \ra (a, \underline{a} \lra c) \\[2pt]
\mathcal R_{3\text{dual}}[t] &: \forall c.\ (\Dual_c[a], \Dual_c[b]) \ra \Dual_c[a]
\end{array} \\
\Dual_c[\R] = (\R, \underline\R \lra c) \qquad
\Dual_c[()] = () \qquad
\Dual_c[(a, b)] = (\Dual_c[a], \Dual_c[b])
\end{gather*}

Their implementations look as follows:
\begin{align*}
\mathcal R_3[t] &= \fun (x : (a, b), \mit{dx} : (\underline{a}, \underline{b}) \lra c).\ (\fst(x), \fun (d : \underline{a}).\ \mit{dx}\ (d, \mbf{0}_{\underline{b}})) \\
\mathcal R_{3\text{dual}}[t] &= \fun (x : (\Dual_c[a], \Dual_c[b])).\ \fst(x)
\end{align*}
where $\mbf{0}_{\underline{b}}$ is the zero value of the cotangent type of $b$.
The issue with the first variant is that $b$ may be an arbitrarily complex type, perhaps even containing large arrays of scalars, and hence this zero value $\mbf{0}_{\underline{b}}$ may also be large.
Having to construct this large zero value is not, in general, possible in constant time, whereas the primal operation ($\fst$) was a constant-time operation; this is anathema to getting a reverse AD code transformation with the correct time complexity.

There are ways to avoid having to construct a non-constant-size zero value here (one of which will be used, in a different situation, later in this paper in \cref{sec:cayley}), but in this paper we choose the $\mathcal R_{3\text{dual}}$ approach.
We do this not to avoid having to deal with the issue of large zeros---indeed, skipping the problem here just moves it somewhere else, namely to the implementation of the backpropagators ($\underline{\R} \lra c$).
However, the $\mathcal R_{3\text{dual}}$ approach extends more easily to a variety of language features (see \cref{sec:source-language-extension}), and more naturally inspires the optimisation to use mutable updates in \cref{sec:mutarrays}, which appears to be necessary to attain the correct complexity for reverse AD on a higher-order language.

\section{Naive, unoptimised dual-numbers reverse AD}\label{sec:naive}


The first AD algorithm described in this paper will be the simplest compositional implementation of the type of $\mathcal R_{3\text{dual}}$: this algorithm is easy to define and prove correct, but is wildly inefficient in terms of complexity.
Indeed, it tends to blow up to exponential overhead over the original function, whereas the desired complexity is to have only a constant factor overhead over the original function.
After the naive algorithm has been discussed, we will apply a number of optimisations to the algorithm (in \cref{sec:staging} and onwards) that fix the complexity issues, in order to finally arrive at an algorithm that does have the desired complexity.

\subsection{Source and target languages}

The reverse AD methods described in this paper are code transformations, and hence have a source language (the language in which input programs may be written) and a target language (in which gradient programs are expressed).
While the source language will be identical for all versions of the transformation that we discuss, the target language will expand to support the optimisations that we perform.

\begin{figure}
  \figheading{Types:}
  \begin{align*}
    \begin{array}{l@{\ \ }c@{\ \ }l}
    \sigma, \tau &\coloneqqq&
      \R \mid () \mid (\sigma, \tau) \mid \sigma \ra \tau \mid \Int
    \end{array}
  \end{align*}
  \figheading{Terms:}
  \begin{align*}
    \begin{array}{l@{\ \ }c@{\ \ }ll}
    s, t &\coloneqqq&
      (x : \tau) & \text{(variable references)} \\
      &\mid& () & \text{(unit constructor)} \\
      &\mid& (s, t) & \text{(pair constructor)} \\
      &\mid& \fst(t) \mid \snd(t) & \text{(pair projections)} \\
      &\mid& s\ t & \text{(function application)} \\
      &\mid& (\fun (x : \tau).\ t) & \text{(lambda abstraction)} \\
      &\mid& (\mbf{let}\ x : \tau = s\ \mbf{in}\ t) & \text{(let binding)} \\
      &\mid& r & \text{(literal $\R$ values)} \\
      &\mid& \mit{op}(t_1,\ldots,t_n) & \text{($\mit{op} \in \Op_n$, primitive operation application ($\R^n \ra \R$))} \\
    \end{array}
  \end{align*}
  \caption{\label{fig:source-language}
    The source language of all variants of this paper's reverse AD transformation.
  }
\end{figure}

The source language is defined in \cref{fig:source-language}; the initial target language is given in \cref{fig:target-language-1}.
The typing of the source language is completely standard, so we omit typing rules here.
We assume call-by-value evaluation.
The only part that warrants explanation is the treatment of primitive operations: for all $n \in \Z_{>0}$ we presume the presence of a set $\Op_n$ containing primitive operations of type $\R^n \ra \R$ in the source language.
The program transformation does not care what the contents of $\Op_n$ are, as long as their derivative operators are available in the target language after differentiation; however, we do require that primitive operation arity is bounded: $\exists m.\ \forall n > m.\ \Op_n = \varnothing$.
(For example by having finitely many primitive operators.)


\begin{figure}
  \figheading{Types:}
  \begin{align*}
    \begin{array}{l@{\ \ }c@{\ \ }ll}
    \sigma, \tau &\coloneqqq&
      \color{gray} \R \mid () \mid (\sigma, \tau) \mid \sigma \ra \tau \mid \Int \\
      &\mid& \sigma \lra \tau & \text{(linear functions)}
    \end{array}
  \end{align*}
  \figheading{Terms:}
  \begin{align*}
    \begin{array}{l@{\ \ }c@{\ \ }ll}
    s, t &\coloneqqq&
      \color{gray} \mathrlap{(x : \tau) \mid () \mid (s, t) \mid \fst(t) \mid \snd(t) \mid s\ t} \\
      &\color{gray} \mid& \color{gray} (\fun (x : \tau).\ t) \mid (\mbf{let}\ x : \tau = s\ \mbf{in}\ t) \\
      &\color{gray} \mid& \color{gray} r \mid (\mit{op} : \sigma \ra \tau) \\
      &\mid& (\lfun(z : \tau).\ b) & \text{(linear lambda abstraction ($\tau$ plain-data type))}
    \end{array}
  \end{align*}
  \figheading{Linear function bodies:}
  \begin{align*}
    \begin{array}{l@{\ \ }c@{\ \ }ll}
    b &\coloneqqq&
      () \mid (b, b') \mid \fst(b) \mid \snd(b) & \text{(tupling)} \\
      &\mid& z & \text{(reference to $\lfun$-bound variable)} \\
      &\mid& (x : \sigma \lra \tau)\ b & \text{(linear function application)} \\
      &\mid& \partial_i\mit{op}(x_1,\ldots,x_n)(b) & \text{($\mit{op} \in \Op_n$, $i$'th partial derivative of $\mit{op}$ ($\R^n \ra \R$))} \\
      &\mid& b + b' & \text{(elementwise addition of results)} \\
      &\mid& \underline0 & \text{(zero of result type)}
    \end{array}
  \end{align*}
  \caption{\label{fig:target-language-1}
    The target language of the unoptimised variant of the reverse AD transformation.
    Components that are also in the source language (\cref{fig:source-language}) are set in \textcolor{gray}{grey}.
  }
\end{figure}

In the target language in \cref{fig:target-language-1}, we add linear functions with the type $\sigma \lra \tau$: these functions are linear in the sense of being commutative monoid homomorphisms, meaning that if $f : \sigma \lra \tau$, then $\sem{f}(0) = 0$ and $\sem{f}(x + y) = \sem{f}(x) + \sem{f}(y)$.\footnote{Actually they are also vector space homomorphisms (hence multiplication by a matrix), but this turns out to be unimportant.}
In order to side-step the issue of defining what it means to be a linear function from or to a function space (and because we do not need it), we require for all $\sigma \lra \tau$ types that $\sigma$ and $\tau$ both do not contain function types ($\ra$, $\lra$).\footnote{In \cref{sec:cayley} we will, actually, put endomorphisms ($a \ra a$) on both sides of a $\lra$-arrow; for justification, see there.}
Operationally, however, linear functions are just regular functions: the operational meaning of all code in this paper remains identical if all $\lra$-arrows are replaced with $\ra$ (and partial derivative operations are allowed in regular terms).


On the term level, we add an introduction form for linear functions; because we disallowed linear function types from or to function spaces, neither $\tau$ nor the type of $t$ can contain function types in $(\lfun(z : \tau).\ t)$.
The body of such linear functions is given by the restricted term language under $b$, which adds application of linear functions (identified by a variable reference), partial derivative operators, and zero and plus operations, but removes variable binding and lambda abstraction.
Note that zero and plus will always be of a type which is (part of) the domain or codomain of a linear function, and hence will have the commutative monoid structure required for these operations to make sense.
The fact that these two operations are not constant-time will be addressed when we improve the complexity of our algorithm later.

Regarding the derivatives of primitive operations: in a linear function, we must be able to compute the linear (reverse) derivatives of the primitive operations.
For every $\mit{op} \in \Op_n$ we require the availability of $\partial_i\mit{op} : \R^n \ra (\R \lra \R)$ with the semantics $\sem{\partial_i\mit{op}}(x)(d) = d \cdot \frac{\partial (\sem{\mit{op}}(x))}{\partial x_i}$.
The $b$ argument in the $\partial_i\mit{op}$ form in the target language is a linear function body term with type $\R$, since the codomain of all primitive operations is $\R$.
(The primal arguments to derivative operators will always be variable references, and are hence indicated as such.)

A type system for the source and (initial) target language is given in \cref{app:type-system}.

\subsection{The code transformation}

The naive reverse AD algorithm only acts non-homomorphically over the syntax of the input program at the places where something non-trivial is done with real scalars.
The full program transformation is given in \cref{fig:algo-naive}.

\begin{figure}
  \figheading{On types:}
  \begin{gather*}
    \trans1c[\R] = (\R, \R \lra c) \qquad
    \trans1c[()] = () \qquad
    \trans1c[(\sigma,\tau)] = (\trans1c[\sigma], \trans1c[\tau]) \\
    \trans1c[\sigma \ra \tau] = \trans1c[\sigma] \ra \trans1c[\tau] \qquad
    \trans1c[\Int] = \Int \\
    \text{On environments:} \qquad
      \trans1c[\varepsilon] = \varepsilon \qquad
      \trans1c[\Gamma, x : \tau] = \trans1c[\Gamma], x : \trans1c[\tau]
  \end{gather*}
  \figheading{On terms:}
  \begin{align*}
    &\text{If}\ \Gamma \vdash t : \tau\ \text{then}\ \trans1c[\Gamma] \vdash \trans1c[t] : \trans1c[\tau] \\
    &\textcolor{gray}{\trans1c[(x : \tau)] = (x : \trans1c[\tau])} \\
    &\textcolor{gray}{\trans1c[()] = ()} \\
    &\textcolor{gray}{\trans1c[(s,t)] = (\trans1c[s], \trans1c[t])} \\
    &\textcolor{gray}{\trans1c[\fst(t)] = \fst(\trans1c[t])} \\
    &\textcolor{gray}{\trans1c[\snd(t)] = \snd(\trans1c[t])} \\
    &\textcolor{gray}{\trans1c[s\ t] = \trans1c[s]\ \trans1c[t]} \\
    &\textcolor{gray}{\trans1c[(\fun(x : \tau).\ t)] = (\fun(x : \trans1c[\tau]).\ \trans1c[t])} \\
    &\textcolor{gray}{\trans1c[(\mbf{let}\ x : \tau = s\ \mbf{in}\ t)] = (\mbf{let}\ x : \trans1c[\tau] = \trans1c[s]\ \mbf{in}\ \trans1c[t])} \\
    &\trans1c[r] = (r, \lfun(z : \R).\ \underline0) \\
    &\trans1c[\mit{op}(t_1,\ldots,t_n)] = (\mbf{let}\ (x_1, d_1) = \trans1c[t_1]\ \mbf{in}\ \ldots\ \mbf{in}\ \mbf{let}\ (x_n, d_n) = \trans1c[t_n] \\
    &\hspace{3.10cm} \mbf{in}\ (\mit{op}(x_1, \ldots, x_n) \\
    &\hspace{3.53cm} , \lfun(z : \R).\ d_1\ (\partial_1\mit{op}(x_1, \ldots, x_n)(z)) + \cdots + \\
    &\hspace{5.49cm} d_n\ (\partial_n\mit{op}(x_1, \ldots, x_n)(z))))
  \end{align*}
  \caption{\label{fig:algo-naive}
    The naive code transformation from the source (\cref{fig:source-language}) to the target (\cref{fig:target-language-1}) language.
    The cases where $\trans1c$ just maps homomorphically over the source language are set in \textcolor{gray}{grey}.
  }
\end{figure}

Note that in the transformation, we take some liberties with notation of the target language: we use implicitly destructuring let-bindings, where $\mbf{let}\ (x_1, x_2) = s\ \mbf{in}\ t$ should be read as $\mbf{let}\ y = s\ \mbf{in}\ \mbf{let}\ x_1 = \fst(y)\ \mbf{in}\ \mbf{let}\ x_2 = \snd(y)\ \mbf{in}\ t$, where $y$ is fresh.

The transformation consists of a mapping $\trans1c[\tau]$ on types $\tau$ and a mapping $\trans1c[t]$ on terms $t$.
The mapping on types works homomorphically except on scalars, which it maps (in the style of dual-numbers AD) to a \emph{pair} of a scalar and a derivative of that scalar.
In constrast to forward AD, however, the derivative is not represented by another scalar (which in forward AD would contain the derivative of this scalar result with respect to a particular initial input value), but instead by a \emph{backpropagator} (see \cref{sec:rev-ad-type}) that maps the reverse derivative of this scalar (the partial derivative of the final result with respect to this scalar) to the reverse derivative of the full input, assuming that the result depends only on the input through this scalar.
(We will explain how this proceeds in \cref{sec:wrapper}.)

The mapping on types is extended to a mapping on environments by applying $\trans1c$ to all types in the environment.

For the implementation, let us go over the syntactic constructs of the source language.
Scalar constants are transformed to a pair of that scalar constant and a backpropagator that produces the derivative of the input given the derivative of this subterm.
This input derivative is of course zero, since the current subterm (a scalar constant) does not depend on the input.

Variable references, tuples, projections, function application, lambda abstraction and let-binding are mapped homomorphically, which is to say that the code transformation simply recurses over the subterms of the current term.
However, it is important to note that for variable references, lambda abstraction and let-binding, the types of the variables do change.

Finally, primitive scalar operations are the most important place where this code transformation does something non-trivial.
First we compute the values and backpropagators of the (scalar) arguments to the operation, after which we can return the (scalar) result of the operation as usual by projecting out the arguments from the computed pairs.
Additionally we return the backpropagator of the result of the operation, which applies all partial derivatives of the primitive operation to the incoming cotangent derivative, passes the results on to the corresponding backpropagators of the arguments, and finally adds the (top-level input derivative) results of type $c$ together.

\subsection{Wrapper of the AD transformation}\label{sec:wrapper}

As given, the code transformation in \cref{fig:algo-naive} has a significant downside: the user of the transformation has to supply backpropagators for the input manually, and if the result type $\tau$ is not $\R$ but instead some more complex type, the user furthermore has to call all the backpropagators in the output individually.
Indeed, while the type of $\mathcal R_{3\text{dual}}$ of \cref{sec:rev-ad-type} yields a very straightforward code transformation, $\mathcal R_2$ is more friendly to the user.

\begin{figure}
  \begin{align*}
    &\begin{array}{@{}l@{\ }c@{\ }l}
      \Interleave1_\tau &:& \forall c.\ (\tau, \tau \lra c) \ra \trans1c[\tau] \\
      \Interleave1_{\R} &=& \fun(x, d).\ (x, d) \\
      \Interleave1_{()} &=& \fun((), d).\ () \\
      \Interleave1_{(\sigma, \tau)} &=& \fun((x, y), d).\ (\Interleave1_\sigma\ (x, \lfun(z : \sigma).\ d\ (z, \underline0)) \\
      &&\hspace{1.765cm} ,\hspace{-0.02cm}\Interleave1_\tau\ (y, \lfun(z : \tau).\ d\ (\underline0, z))) \\
      \Interleave1_{\Int} &=& \fun(n, d).\ n \\
      \Interleave1_{\sigma \ra \tau} &=& \text{not defined!}
    \end{array} \\[0.2cm]
    &\begin{array}{@{}l@{\ }c@{\ }l}
      \Deinterleave1_\tau &:& \forall c.\ \trans1c[\tau] \ra (\tau, \tau \lra c) \\
      \Deinterleave1_{\R} &=& \fun(x, d).\ (x, d) \\
      \Deinterleave1_{()} &=& \fun().\ ((), \lfun(z : ()).\ \underline0) \\
      \Deinterleave1_{(\sigma, \tau)} &=& \fun(x, y).\ \mbf{let}\ (x_1, x_2) = \Deinterleave1_\sigma\ x \\
      &&\hspace{1.13cm} \mbf{in}\ \mbf{let}\ (y_1, y_2) = \Deinterleave1_\tau\ y \\
      &&\hspace{1.13cm} \mbf{in}\ ((x_1, y_1), \lfun(z : (\sigma, \tau)).\ x_2\ (\fst(z)) + y_2\ (\snd(z))) \\
      \Deinterleave1_{\Int} &=& \fun n.\ (n, \lfun(z : \Int).\ \underline0) \\
      \Deinterleave1_{\sigma \ra \tau} &=& \text{not defined!}
    \end{array} \\[0.2cm]
    &\begin{array}{@{}l}
      \Wrap1 : (\sigma \ra \tau) \rightsquigarrow (\sigma \ra (\tau, \tau \lra \sigma)) \\
      \Wrap1[\fun(x : \sigma).\ t] = \fun(x : \sigma).\ \mbf{let}\ x : \trans1\sigma[\sigma] = \Interleave1_\sigma\ (x, \id)\ \mbf{in}\ \Deinterleave1_\tau\ (\trans1\sigma[t])
    \end{array}
  \end{align*}
  \caption{\label{fig:wrapper-naive}
    Wrapper around $\trans1c$ of \cref{fig:algo-naive}.
  }
\end{figure}

Hence, in \cref{fig:wrapper-naive},\footnote{Like in \cref{fig:algo-naive}, we take some notational liberties in \cref{fig:wrapper-naive} regarding destructuring bindings: in addition to destructuring let-bindings, one should now also read $\fun(x_1,x_2).\ t$ to be shorthand for $\fun y.\ t[\fst(y)/x_1][\snd(y)/x_2]$ where $y$ is fresh.} we give a wrapper $\Wrap1$ around $\trans1c$ that exposes an interface in the style of $\mathcal R_2$ instead (taking $\underline\tau = \tau$).
The input program (pattern-matched as ``$\fun(x : \sigma).\ t : \tau$'') must be a closed function of type $\sigma \ra \tau$.\footnote{
  We restrict to the closed case here for simplicity, but this is by no means necessary: by adjusting $\Wrap{}$, one has the option of e.g.\ differentiating with respect to the entire environment, or to regard free variables as constants.
}
We also choose $c$ here, the target type of the backpropagators that was still universally quantified up until now: since it must hold the reverse derivative of the top-level input, which is of type $\sigma$, we simply choose $c = \sigma$.
Note that this choice for $c$ is not unexpected, since that was precisely what we did in \cref{sec:rev-ad-type} to convert back from $\mathcal R_3$ to $\mathcal R_2$.

However, our $\trans1c$ is in the style of $\mathcal R_{3\text{dual}}$, not $\mathcal R_3$, and this difference is bridged using $\Interleave1$ and $\Deinterleave1$.
These two functions are defined inductively over the input and output types of the program being differentiated, and make the assumption that neither of these types contains function types.
Note that this is not a harsh restriction in practice, because it is unclear what the meaning of the derivative with respect to a function value would be in the first place.

All ingredients are combined in $\Wrap1$, where $\Interleave1_\sigma$ is used to build the input to $\trans1\sigma[t]$ of type $\trans1\sigma[\sigma]$ from $x : \sigma$ and $\id : \sigma \lra \sigma$, $\trans1\sigma[t]$ then uses the bound $x$ to compute a value of type $\trans1\sigma[\tau]$, and finally $\Deinterleave1_\tau$ converts this result back to a value of type $(\tau, \tau \lra \sigma)$: the usual function result paired with the reverse derivative function.
Note that the full typing of $t$ is $x : \sigma \vdash t : \tau$, and therefore we have $x : \trans1\sigma[\sigma] \vdash \trans1\sigma[t] : \trans1\sigma[\tau]$, meaning that with $x : \trans1\sigma[\sigma]$ bound in scope, the usage of $\trans1\sigma[t]$ is indeed well-scoped.

\paragraph{Modularity}
Note that the compositionality of $\trans1c$ means that this reverse AD algorithm is modular: we can do the work of differentiating pieces of code with $\trans1c$ without knowledge of where this differentiated code will be used (i.e.\ where the wrapper $\Wrap1$ will be used).
This allows the creation and distribution of libraries of differentiated functions in a natural way.
This modularity property is not a given in reverse AD algorithms: if the core program transformation is not as compositional as $\trans1c$, achieving modularity can be more difficult.

\subsection{Complexity of the naive transformation}\label{sec:naive-complexity}

Reverse AD transformations like the one described in this section are well-known to be correct (e.g.~\cite{ad-2020-dualnum-revad-linear-factoring,ad-2021-dual-revad-linear-factoring-pcf,ad-2020-sam-mathieu-matthijs,nunes-2022-dual-numbers}).
However, as given here, it does not at all have the right time complexity.

The derivative program ($\trans1c[t]$) will invoke the linear backpropagator corresponding to a particular scalar value once for each usage of that scalar value as an argument to a primitive operation; this means that any sharing in the original program (that is: a variable being used more than once) results in multiple calls to the same backpropagator in the derivative program.
All non-trivial backpropagators are either provided as input to the derivative program through the top-level argument of type $\trans1c[\sigma]$ or created upon execution of a primitive operation, and in this latter case, the created backpropagator recursively calls other backpropagators created earlier.
Hence, having multiple calls to a backpropagator usually leads to multiple calls of input backpropagators, and this call multiplication stacks: in general, the number of invocations of input backpropagators can increase exponentially over the runtime of the original program.

There is an instructive example of this phenomenon in the following program:
\begin{align*}
  \begin{array}{l@{\ }l}
    t_{\text{bad}} = \fun(x_0 : \R).
      &\mbf{let}\ x_1 = \mit{op}(x_0, x_0) \qquad\qquad\text{\textit{(for any choice of $\mit{op} \in \Op_2$)}} \\
      &\mbf{in}\ \mbf{let}\ x_2 = \mit{op}(x_1, x_1) \\
      &\quad\vdots \\
      &\mbf{in}\ \mbf{let}\ x_n = \mit{op}(x_{n-1}, x_{n-1}) \\
      &\mbf{in}\ x_n
  \end{array}
\end{align*}
Differentiating this, we get:
\begin{align*}
  &\trans1c[t_{\text{bad}}] = \fun(x_0 : (\R, \R \lra c)). \\
  &\qquad \mbf{let}\ x_1 = (\mit{op}(\fst(x_0), \fst(x_0)), \lfun(z : \R).\ \snd(x_0)\ (\partial_1\mit{op}(\fst(x_0), \fst(x_0))(z)) \ + \\
  &\hspace{6.4cm} \snd(x_0)\ (\partial_1\mit{op}(\fst(x_0), \fst(x_0))(z))) \\
  &\qquad\quad\vdots \\
  &\qquad \mbf{in}\ \mbf{let}\ x_n = (\mit{op}(\fst(x_{n-1}), \fst(x_{n-1})), \lfun(z : \R).\ \snd(x_{n-1})\ (\partial_1\mit{op}(\fst(x_{n-1}), \fst(x_{n-1}))(z)) \ + \\
  &\hspace{7.48cm} \snd(x_{n-1})\ (\partial_1\mit{op}(\fst(x_{n-1}), \fst(x_{n-1}))(z))) \\
  &\qquad \mbf{in}\ x_n
\end{align*}
The destructuring let-bindings in the $\trans1c$ transformation of primitive operation calls were inlined using $\fst$/$\snd$ projections here.
As can be seen, invoking $\snd(x_n)$ (the backpropagator of the result, i.e.\ the function that computes the actual gradient) will eventually invoke $\snd(x_0)$ exactly $2^n$ times.

This certainly sounds suboptimal, and it is indeed completely unacceptable because we can do much bettter (and with such run times, using forward AD to compute a gradient would probably be more efficient): for first-order programs we understand well how to write a code transformation such that the output program computes the gradient in only a constant factor overhead over the original program~\cite{adbook-2008-griewank-walther}.
This is less immediately clear for higher-order programs, as we consider here, but it is nevertheless possible.

In~\cite{ad-2020-dualnum-revad-linear-factoring}, the problem of exponential complexity is addressed by observing that calling a linear backpropagator multiple times is a waste of work: indeed, the very fact that the backpropagators are linear means that, for such a backpropagator $f$, we have $f\ x + f\ y = f\ (x + y)$.
Hopefully, applying this \emph{linear factoring rule} from left to right would allow us to somehow ensure that every backpropagator is executed at most once.
And indeed, if we achieve this, the complexity issue described above is rectified: first, note that the runtime of the forward pass clearly has a constant factor overhead over the original program (similarly to dual-numbers forward AD).
The reverse pass is started by calling the top-level backpropagator returned by $\Deinterleave1_\sigma$, which then starts a call tree that eventually calls all of the backpropagators created in the forward pass.
If we can somehow ensure that every created backpropagator will be executed at most once, we indeed see that the reverse pass would perform work at most proportional to the forward pass for this program: every created backpropagator corresponds to a forward-pass primitive operation invocation, and the amount of work performed inside one single backpropagator (not counting its callees) is constant.\footnote{This is because $\underline0$ and $(+)$ do not yet give significant problems here; we discuss this in more detail in \cref{sec:staged-complexity}.}
For this program, then, the complexity is then as it should be.

However, this argument crucially depends on us being able to ensure that every backpropagator gets invoked at most once.
The solution of~\cite{ad-2020-dualnum-revad-linear-factoring} is to symbolically evaluate the output program of the transformation to a straight-line program with the input backpropagators still as symbolic variables, and afterwards symbolically reduce the obtained straight-line program in a very specific way, making use of the linear factoring rule ($f\ x + f\ y = f\ (x + y)$) in judicious places.
After this reduction, the gradient can be read off from the final, reduced program, which will be in the form $\sum_i (x_i^*\ d_i)$, where $x_i^*$ is the backpropagator of the $i$'th input scalar and $d_i$ is the element of the gradient corresponding to that scalar.

In the (non-standard) variant of the lambda calculus that they use, the above symbolic evaluation procedure results in a number of reduction steps that is proportional to the size of the straight-line program obtained in the process, which should be proportional to the runtime of the original, untransformed program.
However, it is not clear whether the reduction order they suggest can be implemented in an interpreter for the standard lambda calculus, making it unclear whether the complexity guarantees of their algorithm can be attained in a practical implementation.

In this paper, we present an alternative way to use the linear factoring rule to make standard, call-by-value evaluation of the target language have the correct computational complexity, namely by changing the type $c$ that the input backpropagators map to, into something more intelligent than a simple product of cotangent derivatives of all input values.

\section{Linear factoring by staging function calls}\label{sec:staging}

If we take the target type $c$ of the input backpropagators to be equal to the input type $\sigma$ of the original program (of type $\sigma \ra \tau$), as we do in $\Wrap1$ in \cref{fig:wrapper-naive}, all linear functions in the derivative program (as produced by $\trans1\sigma$ from \cref{fig:algo-naive}) have one of the following three forms:
\begin{enumerate}
\item
  $(\lfun(z : \R).\ t)$ where $t$ is a tuple (of type $\sigma$) filled with zero scalars except for one position, where it places $z$; we call such tuples \emph{one-hot tuples}.
  These backpropagators result from $\Interleave1_\sigma$ (\cref{fig:wrapper-naive}) after trivial beta-reduction of the intermediate linear functions.
\item
  $(\lfun(z : \R).\ \underline0)$ occurs as the backpropagator of a scalar constant $r$.
  Note that since this $\underline0$ is of type $\sigma$, operationally it is equivalent to a tuple filled completely with zero scalars.
\item
  $(\lfun(z : \R).\ d_1\ (\partial_1\mit{op}(x_1,\ldots,x_n)(z)) + \cdots + d_n\ (\partial_n\mit{op}(x_1,\ldots,x_n)(z)))$ for an $\mit{op} \in \Op_n$ where $d_1,\ldots,d_n$ are other linear backpropagators: these occur as the backpropagator generated for an application of a primitive operation.
\end{enumerate}
Consequently, all backpropagators contained in the return value (of type $\trans1\sigma[\tau]$) of the derivative program also have one of these three forms.

As observed above in \cref{sec:naive-complexity}, the reverse AD algorithm obtains the correct complexity if we somehow manage to ensure that all backpropagators are invoked at most once, and for that we must use the fact that for a linear function $f$ we have $f\ x + f\ y = f\ (x + y)$.
This means that we must find a way to ``merge'' all invocations of a single backpropagator using this linear factoring rule so that in the end only one invocation remains (or zero if it was not invoked at all in the first place).

\paragraph{Evaluation order}
Ensuring the above-mentioned complete merging is really a question of choosing an order of evaluation for the tree of function calls created by the backpropagators.
Consider for example the (typical) situation where a program generates the following backpropagators:
\begin{align*}
  f_1 &= \fun(z : \R).\ (0, (z, 0)) \\
  f_2 &= \fun(z : \R).\ f_1\ (2 \cdot z) + f_1\ (3 \cdot z) \\
  f_3 &= \fun(z : \R).\ f_2\ (4 \cdot z) + f_1\ (5 \cdot z) \\
  f_4 &= \fun(z : \R).\ f_2\ z + f_3\ (2 \cdot z)
\end{align*}
and where $f_4$ is the (only) backpropagator contained in the result.
With normal call-by-value evaluation, calling $f_4$ first invokes $f_2$ (which in turn calls $f_1$ twice) and then invokes $f_3$ (which in turn calls $f_2$ and $f_1$, for a total of three $f_1$ calls).
$f_2$ was invoked twice and $f_1$ was invoked four times in total.

However, taking inspiration from symbolic evaluation and moving away from standard call-by-value for a moment, we could also first invoke $f_3$ to expand the body of $f_4$ to $f_2\ z + f_2\ (4 \cdot (2 \cdot z)) + f_1\ (5 \cdot (2 \cdot z))$.
Now we can take the two invocations of $f_2$ together using linear factoring to produce $f_2\ (z + 4 \cdot (2 \cdot z)) + f_1\ (5 \cdot (2 \cdot z))$; then invoking $f_2$ first, producing two more calls to $f_1$, we are left with three calls to $f_1$ which we can take together to a single call using linear factoring, which we can then evaluate.
With this alternate evaluation order, we have indeed ensured that every linear function is invoked at most (in this case, exactly) once.

If we want to obtain something like this evaluation order in normal conditions, the first thing we must achieve is to \emph{postpone} invocation of linear functions until we conclude that we have merged all calls to that function, and its time for evaluation has arrived.
To achieve this goal, we will change the representation of $c$ to a dictionary mapping linear functions to their arguments.\footnote{A dictionary with \emph{functions} as keys is suspect, and indeed cannot be implemented in practice. This will be fixed later.}
The idea is that we replace what are now applications of linear functions to creation of a dictionary containing one key-value (function-argument) pair, and to replace addition of values in $c$ by taking the union of dictionaries, where arguments for common keys are added together.

\begin{figure}
  \figheading{On types:}
  \begin{gather*}
    \trans2c[\R] = (\R, \R \lra \Staged c) \qquad
    \textcolor{gray}{\trans2c[()] = ()} \qquad
    \textcolor{gray}{\trans2c[(\sigma,\tau)] = (\trans2c[\sigma], \trans2c[\tau])} \\
    \textcolor{gray}{\trans2c[\sigma \ra \tau] = \trans2c[\sigma] \ra \trans2c[\tau]} \qquad
    \textcolor{gray}{\trans2c[\Int] = \Int}
  \end{gather*}
  \figheading{On terms:}
  \begin{align*}
    &\textcolor{gray}{\text{If}\ \Gamma \vdash t : \tau\ \text{then}\ \trans2c[\Gamma] \vdash \trans2c[t] : \trans2c[\tau]} \\
    &\trans2c[r] = \textcolor{gray}{(r, \lfun(z : \R).}\ \ZeroStaged\textcolor{gray}{)} \\
    &\trans2c[\mit{op}(t_1,\ldots,t_n)] = \textcolor{gray}{(\mbf{let}\ (x_1, d_1) = \trans2c[t_1]\ \mbf{in}\ \ldots\ \mbf{in}\ \mbf{let}\ (x_n, d_n) = \trans2c[t_n]} \\
    &\hspace{3.10cm} \textcolor{gray}{\mbf{in}\ (\mit{op}(x_1, \ldots, x_n)} \\
    &\hspace{3.53cm} \textcolor{gray}{, \lfun(z : \R).}\ \StagedCall\ d_1\ (\partial_1\mit{op}(x_1, \ldots, x_n)(z)) \PlusStaged \cdots \PlusStaged \\
    &\hspace{5.49cm} \StagedCall\ d_n\ (\partial_n\mit{op}(x_1, \ldots, x_n)(z))\textcolor{gray}{))}
  \end{align*}
  \figheading{Updated wrapper:}
  \begin{align*}
    &\begin{array}{@{}l@{\ }c@{\ }l}
      \Interleave2_\tau &:& \forall c.\ (\tau, \tau \lra \Staged c) \ra \trans2c[\tau] \\
      \mathrlap{\text{(Same as \cref{fig:wrapper-naive})}} \\
    \end{array} \\
    &\begin{array}{@{}l@{\ }c@{\ }l}
      \Deinterleave2_\tau &:& \forall c.\ \trans2c[\tau] \ra (\tau, \tau \lra \Staged c) \\
      \mathrlap{\text{($\underline0$ replaced with $\ZeroStaged$, $+$ replaced with $\PlusStaged$)}}
    \end{array} \\
    &\begin{array}{@{}l}
      \Wrap2 : (\sigma \ra \tau) \rightsquigarrow (\sigma \ra (\tau, \tau \lra \sigma)) \\
      \Wrap2[\fun(x : \sigma).\ t] = \fun(x : \sigma).\ \mbf{let}\ x : \trans2\sigma[\sigma] = \Interleave2_\sigma\ (x, \InitStaged) \\
      \hspace{4.405cm} \mbf{in}\ \mbf{let}\ (y, d) = \Deinterleave2_\tau\ (\trans2\sigma[t]) \\
      \hspace{4.405cm} \mbf{in}\ (y, \lfun(z : \tau).\ \ResolveStaged\ (d\ z))
    \end{array}
  \end{align*}
  \figheading{Interface of $\Staged$:}
  \begin{align*}
    \begin{array}{@{}l@{\ }c@{\ }l}
      \ZeroStaged &:& \Staged c \\
      (\PlusStaged) &:& \Staged c \lra \Staged c \lra \Staged c \\
      \StagedCall &:& (\R \lra \Staged c) \ra \R \lra \Staged c \\
      \InitStaged &:& c \lra \Staged c \\
      \ResolveStaged &:& \Staged c \lra c
    \end{array}
  \end{align*}
  \caption{\label{fig:algo-staged}
    The code transformation (from \cref{fig:source-language} to \cref{fig:target-language-1} plus $\Staged$ operations) after staging calls to linear functions as described in \cref{sec:staging}.
    Omitted or grey parts are the same as in \cref{fig:algo-naive}.
  }
\end{figure}

The updated transformation is shown in \cref{fig:algo-staged}.
On the type level we change $\R \lra c$ in the type mapping of $\R$ to $\R \lra \Staged c$; on the term level, in the transformation of a real constant $r$, we replace the $\underline0$ of type $c$ with an explicit zero $\Staged$ object, and in a primitive operation backpropagator we replace the calls to argument backpropagators with staged calls, added together using $(\PlusStaged)$.
We continue to use the notation $\trans2c[\Gamma]$ to map $\trans2c$ over $\Gamma$ like we did in \cref{fig:algo-naive}.

With the updated type of backpropagators for scalars, we also have to make (minor) changes to the wrapper: in some sense, only the types change, but this does mean that the zero and plus of $c$ that occurred in $\Deinterleave1$ need to be replaced with those of $\Staged c$.
In $\Wrap2$, we provide $\Interleave2$ with the initial ($\InitStaged : \sigma \lra \Staged \sigma$) top-level backpropagator instead of the identity function $\sigma \lra \sigma$, and we use $\ResolveStaged$ in the top-level output derivative function to get the final $\sigma$ out.

At this point it is instructive to consider an example, to see how the definitions work together to create a coherent program transformation.
After the example we will give an implementation of the interface of $\Staged$.
For an intuition on the computational cost of the $\Staged c$ interface, for now assume that $\ZeroStaged$, $(\PlusStaged)$, $\StagedCall$ and $\InitStaged$ are\footnote{Actually we will see that all of these are very much \emph{not} constant-time, but not to worry, because this will be fixed later in \cref{sec:cayley}.} $O(1)$, and $\ResolveStaged$ performs $O(n \log n)$ operations in addition to evaluating each of the staged calls exactly once (if they were ever called in the first place; if not, zero times).

\subsection{Example}

Consider the following example program: 
\newcommand\tex{t_{\text{ex}}}
\begin{align*}
  \begin{array}{l}
    x : (\R, \R) \vdash \tex : \R \\
    \tex = \mbf{let}\ x_1 = \fst(x)\ \mbf{in}\ \mbf{let}\ x_2 = \snd(x)\ \mbf{in}\ \mbf{let}\ x_3 = x_1 + x_2\ \mbf{in}\ x_3 \cdot x_1 + x_3
  \end{array}
\end{align*}
Then $(\fun(x : (\R, \R)).\ \tex)$ is a closed function of type $(\R, \R) \ra \R$ and thus a suitable input for $\Wrap2$.
The differentiated target program looks as follows:
\begin{align*}
  \begin{array}{l}
    x : ((\R, \R \lra \Staged{}(\R, \R)), (\R, \R \lra \Staged{}(\R, \R))) \vdash \trans2{(\R,\R)}[\tex] : (\R, \R \lra \Staged{}(\R, \R)) \\[2pt]
    \trans2{(\R,\R)}[\tex] = \mbf{let}\ x_1 = \fst(x)\ \mbf{in}\ \mbf{let}\ x_2 = \snd(x) \\[2pt]
    \hspace{1.958cm} \mbf{in}\ \mbf{let}\ x_3 = (\fst(x_1) + \fst(x_2) \\[2pt]
    \hspace{3.618cm} , \lfun(z : \R).\ \StagedCall\ (\snd(x_1))\ z \PlusStaged \StagedCall\ (\snd(x_2))\ z) \\[2pt]
    \hspace{1.958cm} \mbf{in}\ \mbf{let}\ y_1 = (\fst(x_3) \cdot \fst(x_1) \\[2pt]
    \hspace{3.618cm} , \lfun(z : \R).\ \StagedCall\ (\snd(x_3))\ (\fst(x_1) \cdot z) \PlusStaged {} \\[2pt]
    \hspace{5.498cm} \StagedCall\ (\snd(x_1))\ (\fst(x_3) \cdot z)) \\[2pt]
    \hspace{2.368cm} \mbf{in}\ \mbf{let}\ y_2 = x_3 \\[2pt]
    \hspace{2.368cm} \mbf{in}\ (\fst(y_1) + \fst(y_2) \\[2pt]
    \hspace{2.798cm} , \lfun(z : \R).\ \StagedCall\ (\snd(y_1))\ z \PlusStaged \StagedCall\ (\snd(y_2))\ z)
  \end{array}
\end{align*}
The only difference between the code shown here and the code produced by diligently applying the transformation in \cref{fig:algo-staged} is realisation of the let-bindings in the case for primitive operations as $\fst$ and $\snd$ projections, and as $y_1$ and $y_2$ for the final addition.\footnote{These changes, as well as the $\beta$-reductions in the wrapper directly below, are just for readability and do not influence complexity.}
The wrapper looks as follows:
\begin{align*}
  &\Interleave2_{(\R,\R)}
    = \fun((x,y),d).\ (\bigl(\fun(x,d).\ (x, d)\bigr)\ (x, \lfun(z : \R).\ d\ (z, \underline0)) \\[-2pt]
    &\hspace{4.22cm} , \bigl(\fun(x,d).\ (x, d)\bigr)\ (y, \lfun(z : \R).\ d\ (\underline0, z))) \\
    &\hspace{2cm} \overset{\smash\beta}\ra \fun((x,y),d).\ ((x, \lfun(z : \R).\ d\ (z, \underline0)) \\[-2pt]
    &\hspace{4.3cm} , (y, \lfun(z : \R).\ d\ (\underline0, z))) \\
  &\Deinterleave2_\R = \fun(x,d).\ (x,d) \\
  &\Wrap2[\fun(x : (\R, \R)).\ \tex] = \fun(x : (\R,\R)). \\[-4pt]
  &\hspace{4.4cm} \begin{array}{l}
    \mbf{let}\ x : \trans2\sigma[\sigma] = \Interleave2_{(\R,\R)}\ (x, \InitStaged) \\
    \mbf{in}\ \mbf{let}\ (y, d) = \Deinterleave2_\R\ (\trans2{(\R,\R)}[\tex]) \\
    \mbf{in}\ (y, \lfun(z : \R).\ \ResolveStaged\ (d\ z))
  \end{array} \\
  &\hspace{3.51cm} \overset{\smash\beta}\ra \fun((x,y) : (\R,\R)). \\[-4pt]
  &\hspace{4.4cm} \begin{array}{l}
    \mbf{let}\ x : \trans2\sigma[\sigma] = ((x, \lfun(z : \R).\ \InitStaged\ (z, \underline0)) \\
    \hspace{2.31cm} , (y, \lfun(z : \R).\ \InitStaged\ (\underline0, z))) \\
    \mbf{in}\ \mbf{let}\ (y',d) = \trans2{(\R,\R)}[\tex] \\
    \mbf{in}\ (y', \lfun(z : \R).\ \ResolveStaged\ (d\ z))
  \end{array}
\end{align*}
Skipping some of the bookkeeping involved in (de)interleaving and focusing only on the (slightly) simplified bottommost version of $\Wrap2[\tex]$, we see that the input (bound to $x$ in the first $\mbf{let}$) to the differentiated program $\trans2{(\R,\R)}[\tex]$ is a pair of ``dual-number'' scalars, each of which contain the original input scalar as well as a backpropagator.
This backpropagator is a linear function that, upon invocation, creates a full $\sigma$ (here $(\R,\R)$) value with the argument placed at the position of the input scalar in question, and with all other places zero; we put this ``one-hot'' $\sigma$ value in a $\Staged \sigma$ object using $\InitStaged$ because all our backpropagators now map into $\Staged \sigma$.
(Recall that we want to stage calls to other linear functions, and every linear function has the potential to call other linear functions; the fact that this one directly returns a value leads to the usage of $\InitStaged$.)
These linear functions that create a one-hot $\sigma$ value will sometimes be called ``injector functions'' in this paper.

The differentiated program then uses the interleaved input and computes the result (bound to $y'$, renamed from $y$ for clarity) and a backpropagator for the result (bound to $d$).
For the final return value, then, the primal component is simply $y'$ as computed by the program; the reverse derivative function of type $\tau \lra \sigma$, here $\R \lra (\R,\R)$, applies $d$ to the incoming result cotangent $z$ and then resolves the $\Staged{}(\R,\R)$ to a value of type $(\R,\R)$ using $\ResolveStaged$.

\subsection[The Staged interface]{The $\Staged$ interface}\label{sec:staged-interface}

The intended implementation of $\Staged c$ is a pair of a value of type $c$ and a persistent tree-map that supports purely-functional logarithmic insertion, query, deletion and maximal key query.
The map will be used for staging linear function calls, and hence should map linear functions to their collected arguments.
When two $\Staged$ objects are added using $(\PlusStaged)$, we add the $c$ constants and take the union of their constituent maps, where values for common keys are added together: \emph{this is the precise place where we apply the linear factoring rule $f\ x + f\ y \rightsquigarrow f\ (x + y)$}.
In other words, we would like the interface methods of $\Staged$ to be implemented as shown under the ``Intended implementation'' header in \cref{fig:staged-tree-impl}.

\begin{figure}
  \figheading{Intended implementation}
  \begin{align*}
    &\Staged c = (c, \Map\ (\R \lra \Staged c)\ \R) \\
    &\quad\text{semantics: } (c, \{f_1 \mapsto a_1,\ldots,f_n \mapsto a_n\}) \text{ is equivalent to } c + {\textstyle\sum_{i=1}^n f_i(a_i)} \\
    &\begin{array}{@{\,}l@{\ }c@{\ }ll}
      \ZeroStaged &=& (\underline0, \EmptyMap) & \textit{(empty tree-map)} \\
      (c_1, m_1) \PlusStaged (c_2, m_2) &=& (c_1 + c_2, m_1 \cup_+ m_2) & \textit{(map union; values for equal keys are $+$'d)} \\
      \StagedCall\ f\ x &=& (\underline0, \{f \mapsto x\}) & \textit{(singleton map with one key-value pair)} \\
      \InitStaged\ c &=& (c, \EmptyMap) \\
      \ResolveStaged\ (c, m) &=& ...?
    \end{array}
  \end{align*}
  \figheading{Actual implementation}
  \newcommand\openparen{(} 
  \begin{align*}
    &\Staged c = (c, \Map\ \Int\ (\R \lra \Staged c, \R)) \\
    &\quad\text{semantics: } (c, \{i_1 \mapsto (f_1, a_1),\ldots,i_n \mapsto (f_n, a_n)\}) \text{ is equivalent to } c + {\textstyle\sum_{j=1}^n f_j(a_j)} \\
    &\begin{array}{@{\,}l@{\ }c@{\ }ll}
      \ZeroStaged &=& (\underline0, \EmptyMap) \\[1mm]
      (c_1, m_1) \PlusStaged (c_2, m_2) &=& (c_1 + c_2, m_1 \cup_{(\fun(f,a)\,(g,b).\;(f,a+b))} m_2) \\
        && \mathrlap{\qquad \textit{\openparen in this map union, values for equal keys are}} \\
        && \mathrlap{\qquad\qquad \textit{combined using $(\fun(f,a)\,(g,b).\;(f,a+b))$.)}} \\[1mm]
      \StagedCall &:& (\Int, \R \lra \Staged c) \ra \R \lra \Staged c & \quad\textit{(note, changed type!)} \\
      \StagedCall\ (i, f)\ x &=& (\underline0, \{i \mapsto (f, x)\}) \\[1mm]
      \InitStaged\ c &=& (c, \EmptyMap) \\[1mm]
      \ResolveStaged\ (c, m) &=& ...? & \hspace{-2.3cm}\textit{(we will return to $\ResolveStaged$ later)}
    \end{array}
  \end{align*}
  \caption{\label{fig:staged-tree-impl}
    The tree-map-based implementation of $\Staged$; \cref{fig:algo-staged} uses the ``Intended implementation'', whereas \cref{fig:algo-monadic} will use the ``Actual implementation''.
  }
\end{figure}

However, as foreshadowed by the ``Actual implementation'' part of \cref{fig:staged-tree-impl}, there is a problem with the intended implementation: one cannot reasonably create a tree map with linear functions as keys, because for that one would need a total order on linear functions, which is impossible in general.\footnote{Hence, $\trans2c$ and $\Wrap2$ cannot actually be implemented.}
However, we can make such a total order \emph{on the functions that we create using $\lfun$} by tagging each such function at runtime with a unique integer!
For the implementation of $\ResolveStaged$, which we discuss below, we will further require (and ensure) that these tags are \emph{monotonically increasing}: a linear function created later must have a tag that is greater than those of functions created earlier in call-by-value ordering.\footnote{%
  For this temporal relation, the order of evaluation for (primitive) functions with multiple arguments (in other words, the degree of freedom in evaluation order that remains after choosing for call-by-value) does not actually matter: the only actual, underlying requirements of the algorithm are that the tags are unique and that a linear function $f$ that in its closure refers to another linear function $g$ must have a tag greater than the tag of $g$.
  To ensure this, it is sufficient to make tags monotonically increasing in any particular call-by-value order.
}
Generating these monotonically increasing tags makes the program produced by the code transformation monadic in an ID generation monad (e.g.\ implementable using a state monad).
This has the downside of forcibly sequentialising the entire computation if it was not already sequential before; see \cref{sec:future-work} for some notes on this.

In the ``Actual implementation'' part of \cref{fig:staged-tree-impl}, two things changed: firstly, the $\Map$ now has integers as keys instead of functions, and secondly, the type of $\StagedCall$ changed to take the integer tag of the called linear function as an additional argument (in a pair for convenience in $\trans3c[\mit{op}(\ldots)]$).
The algorithm assumes that one linear function (i.e.\ the closure object resulting from one invocation of a $\lfun$) is always consistently tagged with the same integer.

\begin{figure}
  \figheading{On types:}
  \begin{gather*}
    \trans3c[\R] = (\R, \standout{(\Int, \R \lra \Staged c)}) \qquad
    \textcolor{gray}{\trans3c[()] = ()} \qquad
    \textcolor{gray}{\trans3c[(\sigma,\tau)] = (\trans3c[\sigma], \trans3c[\tau])} \\
    \trans3c[\sigma \ra \tau] = \trans3c[\sigma] \ra \standout{\Int \ra (\trans3c[\tau], \Int)} \qquad
    \textcolor{gray}{\trans3c[\Int] = \Int}
  \end{gather*}
  \figheading{On terms:}
  \begin{align*}
    &\text{If}\ \Gamma \vdash t : \tau\ \text{then}\ \trans3c[\Gamma] \vdash \trans3c[t] : \standout{\Int \ra (\trans3c[\tau], \Int)} \\
    &\trans3c[(x : \tau)] = \fun i.\ ((x : \trans3c[\tau]), i) \\
    &\trans3c[(s,t)] = \fun i.\ \mbf{let}\ (x, i') = \trans3c[s]\ i\ \mbf{in}\ \mbf{let}\ (y, i'') = \trans3c[t]\ i'\ \mbf{in}\ ((x, y), i'') \\
    &\trans3c[(\mbf{let}\ x : \tau = s\ \mbf{in}\ t)] = \fun i.\ \mbf{let}\ (x : \trans3c[\tau], i') = \trans3c[s]\ i\ \mbf{in}\ \trans3c[t]\ i' \\
    &\textit{etc.\ (standard monadically lifted code)} \\
    &\trans3c[r] = \fun i.\ ((r, (\standout{i}, \lfun(z : \R).\ \ZeroStaged)), \standout{i + 1}) \\
    &\trans3c[\mit{op}(t_1,\ldots,t_n)] = \fun i.\ (\mbf{let}\ ((x_1, d_1), i_1) = \trans3c[t_1]\ i\ \mbf{in}\ \ldots\ \mbf{in}\ \mbf{let}\ ((x_n, d_n), i_n) = \trans3c[t_n]\ i_{n-1} \\
    &\hspace{3.57cm} \mbf{in}\ ((\mit{op}(x_1, \ldots, x_n) \\
    &\hspace{4.15cm} , (\standout{i_n} \\
    &\hspace{4.33cm} , \lfun(z : \R).\ \StagedCall\ d_1\ (\partial_1\mit{op}(x_1, \ldots, x_n)(z)) \PlusStaged \cdots \PlusStaged \\
    &\hspace{6.14cm} \StagedCall\ d_n\ (\partial_n\mit{op}(x_1, \ldots, x_n)(z)))) \\
    &\hspace{4.01cm} , \standout{i_n + 1}))
  \end{align*}
  \figheading{Adjusted wrapper:}
  \begin{align*}
    &\begin{array}{@{}l}
      \Wrap3 : (\sigma \ra \tau) \rightsquigarrow (\sigma \ra (\tau, \tau \lra \sigma)) \\
      \Wrap3[\fun(x : \sigma).\ t] = \fun(x : \sigma).\ \mbf{let}\ (x : \trans3\sigma[\sigma], i) = \Interleave3_\sigma\ (x, \InitStaged)\ 0 \\
      \hspace{4.412cm} \mbf{in}\ \mbf{let}\ (y, d) = \Deinterleave3_\tau\ (\fst(\trans3\sigma[t]\ i)) \\
      \hspace{4.412cm} \mbf{in}\ (y, \lfun(z : \tau).\ \ResolveStaged\ (d\ z))
    \end{array} \\
    &\begin{array}{@{}l@{\ }c@{\ }l}
      \Interleave3_\tau &:& \forall c.\ (\tau, \tau \lra \Staged c) \ra \standout{\Int \ra (\trans3c[\tau], \Int)} \\
      \Interleave3_\R &=& \fun(x, d).\ \fun i.\ ((x, (\standout{i}, d)), \standout{i + 1}) \\
      \Interleave3_{()} &=& \fun((), d).\ \fun i.\ ((), i) \\
      \Interleave3_{(\sigma, \tau)} &=& \fun((x, y), d).\ \fun i.\ \mbf{let}\ (x', i') = \Interleave3_\sigma\ (x, \lfun(z : \sigma).\ d\ (z, \underline0))\ i \\
      &&\hspace{2.19cm} \mbf{in}\ \mbf{let}\ (y', i'') = \Interleave3_\tau\ (y, \lfun(z : \tau).\ d\ (\underline0, z))\ i' \\
      &&\hspace{2.19cm} \mbf{in}\ ((x', y'), i'') \\
      \Interleave3_\Int &=& \fun(n, d).\ \fun i.\ (n, i)
    \end{array} \\
    &\text{$\Deinterleave3_\tau$ gets type $\forall c.\ \trans3c[\tau] \ra (\tau, \tau \lra \Staged c)$ and ignores the new $\Int$ in $\trans3c[\R]$.}
  \end{align*}
  \caption{\label{fig:algo-monadic}
    The monadically transformed code transformation (from \cref{fig:source-language} to \cref{fig:target-language-1} plus $\Staged$ operations), based on \cref{fig:algo-staged}.
    \textstandout{Highlighted} are some places with material changes related to the ID tagging.
  }
\end{figure}

The new $\Staged$ implementation is used in a new monadic code transformation, shown in \cref{fig:algo-monadic}.
On the type level, linear functions (in $\trans3c[\R]$) are now tagged with an integer, and functions in the program now live in the ID generation monad.
On the term level, for most source language constructs (variables, tuples, projections, application, abstraction and let-binding), the output code is simply lifted to monadic code as usual.
We choose to write out the pair administration instead of using the monadic bind notation abstraction to make clear what operations are being executed and where the lambdas occur.
It should be clear to the reader that the only thing that happens here is threading through of the ``next ID to generate'' integer---we understand it as a monad, but operationally it is just threading through the integer.

The only places where actual new IDs are being generated (i.e.\ where the integer is used and incremented) are where linear functions are being created, i.e.\ for scalar constants and in the code for primitive operations.
The output code for primitive operations has a large tuple that might be hard to read; it has shape $((\mit{op}, (i, \lfun)), i + 1)$ and is of type $(\trans3c[\R], \Int)$.

In the wrapper, we must (monotonically) number all input backpropagators in $\Interleave3$ and then execute the monadic action defined by $\trans3\sigma[t]$ with as starting ID the next available ID after the input backpropagators.
Note that in $\Deinterleave3$, and for the final, top-level derivative function, we do create new linear functions that remain untagged; however, this is not an issue because we ensured that all linear functions that are potentially called multiple times (namely the input backpropagators and those created in $\trans3c[\R]$) are in fact tagged.

\paragraph{Resolve order}
We still have to specify how exactly $\ResolveStaged$ is going to ensure that every linear function in a $\Staged$ object is called only once.
The crucial observation here is that in the case of our code transformation, a linear function will only ever call other linear functions that are created earlier during evaluation.
This is why we tag created linear functions with monotonically increasing IDs \emph{at runtime}, because this allows us to read off the dependency structure of linear functions from their tags: a linear function with tag $i$ will only call other linear functions with tags smaller than $i$.

To see that the observation is true for our code transformation, recall that all created backpropagators have, after beta-reduction of the intermediate linear functions in $\Interleave3$, one of three forms: $\lfun(z : \R).\ \ZeroStaged$ for scalar constants, $\lfun(z : \R).\ (\text{one-hot $\sigma$ value with $z$})$ for injectors created in $\Interleave{}$, and the following for primitive operations:
\begin{align*}
&\lfun(z : \R).\ \StagedCall\ d_1\ (\partial_1\mit{op}(x_1,\ldots,x_n)(z)) \PlusStaged \cdots \PlusStaged \\
&\hspace{2cm}\StagedCall\ d_n\ (\partial_n\mit{op}(x_1,\ldots,x_n)(z))
\end{align*}
The only linear functions called (indirectly now, due to staging) by any of these forms of backpropagator are $\snd(d_1),\ldots,\snd(d_n)$ for primitive operations, and by inspecting \cref{fig:algo-monadic} we see that these linear functions are themselves the backpropagators of the arguments of the primitive operation.
And indeed, in call-by-value evaluation\footnote{%
  Even with call-by-need (lazy evaluation), as Haskell implements, we get a call-by-value \emph{numbering} (though not necessarily creation) of the backpropagators, due to the explicit monadic threading of the next ID to generate.
  Hence, the algorithm works fine also with a call-by-need target language (as expected, because lambda calculus evaluation is confluent, and we do not use any non-standard extensions of the standard operational semantics), even though we assume a call-by-value source language.
}, these functions will have been created earlier than the execution of the primitive operation itself.

Now we recall the goal of $\ResolveStaged$: to evaluate the calls to backpropagators one by one in such an order as to ensure that the arguments of all calls to a particular backpropagator are added together before that backpropagator is first invoked.
This then accomplishes the higher goal of evaluating every backpropagator at most once.
In light of the observation above about the dependency order of the backpropagators, we give the following (tail-)recursive definition for $\ResolveStaged$, assuming standard operations on a $\Map$:
\begin{align*}
  \begin{array}{l}
    \ResolveStaged\ (c : \sigma, m : \Map\ \Int\ (\R \lra \Staged \sigma, \R)) \coloneqq \\
    \quad\mbf{if}\ m\ \text{is empty} \\
    \quad\quad\mbf{then}\ c \\
    \quad\quad\mbf{else}\ \mbf{let}\ i = \text{highest key in}\ m \\
    \quad\quad\hphantom{\mbf{else}}\ \mbf{in}\ \mbf{let}\ (f, a) = \text{lookup}\ i\ \text{in}\ m \\
    \quad\quad\hphantom{\mbf{else}}\ \mbf{in}\ \mbf{let}\ m' = \text{delete}\ i\ \text{from}\ m \\
    \quad\quad\hphantom{\mbf{else}}\ \mbf{in}\ \ResolveStaged\ (f\ a \PlusStaged (c, m'))
  \end{array}
\end{align*}
where $(\PlusStaged)$ is (still) defined as in the ``Actual implementation'' in \cref{fig:staged-tree-impl}.
Because a backpropagator can only call other backpropagators with smaller tags, as observed above, this definition will combine (with $(\PlusStaged)$) all staged calls to a particular backpropagator before actually invoking the function, which achieves the goal.
This completes \cref{fig:staged-tree-impl}.

\subsection{Complexity}\label{sec:staged-complexity}

Finally, we look at the complexity of the reverse AD transformation given in \cref{fig:algo-monadic} and indicate what must still be done to fix this complexity.\footnote{This section does not provide a proof that $\Wrap3{}$ does \emph{not} have the correct complexity; rather, it argues that the expected complexity analysis does not go through. The same complexity analysis \emph{will} go through for $\Wrap5$ at the end of \cref{sec:mutarrays}.}

In \cref{sec:rev-ad-complexity}, we gave a property expressing the complexity requirement for a reverse AD code transformation.
With our type signature of $\Wrap{}$, this property translates to the following: (in this section, take $\Wrap{} = \Wrap3$)
\begin{align}
  \begin{array}{l}
    \exists c > 0.\ \forall P : \mrm{Programs}(\sigma \ra \tau).\ \forall x : \sigma, \mit{dy} : \tau. \\[0.7ex]
    \qquad \cost(\snd(\Wrap{}[P]\ x)\ \mit{dy}) \leq c \cdot (\cost(P\ x) + \size(x))
  \end{array}
  \label{eq:complexity-property-wrap}
\end{align}
where we recall that $\cost(E)$ denotes the time taken to evaluate $E$ to normal form and $\size(x)$ denotes the time taken to read all of $x$ sequentially.
We assume without loss of generality that $P$ is of the form ($\fun(x : \sigma).\ t$).
Then, since $P$ and $\Wrap{}[P]$ have a lambda abstraction at the top level, they are already in normal form and hence take constant cost to reduce.

We will analyse this left-hand side in three parts:
\begin{enumerate}
\item\label{item:complexity-part-primal}
  The cost of evaluating $\trans3\sigma[t]\ i$ with the interleaved input in the environment;
\item\label{item:complexity-part-wrapper}
  The cost of calling $\Wrap3[P]$ with argument $x$, which includes (\ref{item:complexity-part-primal}); and
\item\label{item:complexity-part-dual}
  The cost of invoking the top-level backpropagator $\snd(\Wrap3[P]\ x)$ on a cotangent value.
\end{enumerate}
Since part (\ref{item:complexity-part-wrapper}) plus (\ref{item:complexity-part-dual}) is the left-hand side of \cref{eq:complexity-property-wrap}, it is sufficient if we can show that each of these parts runs in time $O(\cost(P\ x) + \size(x))$.

\paragraph{Primal computation time (\ref{item:complexity-part-primal})}
It is easy to see in \cref{fig:algo-monadic} that the work performed by running the monadic action $\trans3c[t]$ for some term $t$ (i.e.\ calling $\trans3c[t]$ with an integer argument) is proportional to the original runtime of $t$.
Note that this does \emph{not} include calling the backpropagators, but it does involve \emph{creating} the backpropagators, which takes time proportional to the size of their closure.
Since in our case the closures are of bounded size (since primitive operators have bounded arity), part (\ref{item:complexity-part-primal}) runs in $O(\cost(P\ x))$, which is stricter than $O(\cost(P\ x) + \size(x))$.

\paragraph{Wrapper computation time (\ref{item:complexity-part-wrapper})}
$\Wrap3$ consists of interleaving, calling the transformed program, and deinterleaving.
The interleaving process itself runs in $O(\size(x))$; the content of the created backpropagators is problematic (as we will see below), but just creating the backpropagators is fine because their closures are small.
Calling the transformed program takes $O(\cost(P\ x))$ as discussed above.
Finally, $\Deinterleave3$, which is similar in structure to $\Deinterleave1$ from \cref{fig:wrapper-naive}, completes in time $O(\size(P\ x))$, i.e.\ the size of the original output.
This is in particular also in $O(\cost(P\ x))$.
Hence, $\cost(\Wrap3[P]\ x)$ is in $O(\cost(P\ x) + \size(x))$ as required.

\paragraph{Dual computation time (\ref{item:complexity-part-dual})}
The reverse derivative propagation pass is in a sense the core of reverse AD, and as expected it is also the most involved part to analyse and make efficient.

The top-level backpropagator $\snd(\Wrap3[P]\ x)$ first calls the backpropagator $d$ returned by $\Deinterleave3_\tau$ to obtain an initial $\Staged \sigma$, and then (in $\ResolveStaged$) calls all created backpropagators at most once.
Furthermore, $\ResolveStaged$ performs some $\Map$ operations per invoked backpropagator, and uses $(\PlusStaged)$ once per invoked backpropagator.

The backpropagator $d$ returned by $\Deinterleave3_\tau$ calls all backpropagators directly contained in the output of the transformed code exactly once, and combines their results using $(\PlusStaged)$.
\begin{itemize}
\item
  \underline{Problem}: Since $(\PlusStaged)$ can be called here as many times as the number of scalars in the program output (minus one), and the program output size is potentially in $O(\cost(P\ x))$, $(\PlusStaged)$ must be constant-time lest we overshoot the complexity budget of $O(\cost(P\ x) + \size(x))$ that we set ourselves in this analysis.
  However, $(\PlusStaged)$ adds $c$ values (which can be large---indeed, that is the whole point of choosing reverse AD over forward AD) and combines staging maps, neither of which are constant-time (because these staging maps are not necessarily of bounded size).
\end{itemize}
Looking past this problem, $\ResolveStaged$ then uses the staging map returned by $d$ to call every created backpropagator at most once.
The number of created backpropagators is equal to the number of times a scalar constant is created or a primitive operator is executed in the original program (these are created by $\trans3c$) plus the number of scalars in the input (these are created by $\Interleave3$).
Hence, this number is in $O(\cost(P\ x) + \size(x))$, meaning that the body of each backpropagator must be constant-time for our complexity analysis to go through.

However, the backpropagator bodies are unfortunately not constant-time.
\begin{itemize}
\item
  \underline{Problem}: The backpropagators created in $\Interleave3_\sigma$ are too expensive.
  Due to the build-up in $\Interleave3_{(\sigma,\tau)}$, the backpropagators joined to individual scalars end up being of the following form:
  \[ \lfun(z : \R).\ \InitStaged\ (0, \ldots, 0, z, 0, \ldots, 0) \]
  if we flatten all pairs into a single large tuple.
  This backpropagator takes time $O(\size(x))$ to run, which is not constant.
\item
  \underline{Problem}: The backpropagators for scalar constants and primitive operations are too expensive.
  Indeed, $\ZeroStaged$ and $\StagedCall$ need to create a zero value of type $c$ (the original program input), and $(\PlusStaged)$ needs to add $c$ values (in addition to merging the maps, but these maps are of bounded size, making the map merge cost negligible).
  None of these three operations are constant-time, while they need to be.
  Note that since primitive operation arity is bounded, $\StagedCall$ and $(\PlusStaged)$ being constant-time would indeed imply that the operator backpropagator runs in constant time as well.
\end{itemize}
Finally, $\ResolveStaged$ (as presented above in \cref{sec:staged-interface}) also performs, per backpropagator invocation, a constant number of $\Map$ operations and one $(\PlusStaged)$ invocation.
The $\Map$ operations are logarithmic in $\cost(P\ x)$, while they need to be constant-time because they are already executed $O(\cost(P\ x))$ times; and the usage of $(\PlusStaged)$ is a problem just like before: even though one of the staging maps that is combined, namely $f\ a$, can be shown to be of constant size, the map union still results in a logarithmic overhead, and the addition of $c$ values is in any case still problematic.

In summary, the parts of the transformed program that are still too expensive are the following:
\begin{enumerate}
\item\label{item:problem-zero}
  $\ZeroStaged$ and $\StagedCall$ must be constant-time, but are not due to the creation of a zero of type $c$.
\item\label{item:problem-plus}
  $(\PlusStaged)$ must be constant-time, but is not due to adding values of type $c$ and taking the union of staging maps.
\item\label{item:problem-onehot}
  The injector backpropagators created by $\Interleave3$ must be constant-time, but are not due to the creation of a one-hot (i.e.\ mostly zero) value of type $c$.
\item\label{item:problem-map}
  The logarithmic $\Map$ operations in $\ResolveStaged$ should be constant-time.
\end{enumerate}
In the next step (\cref{sec:cayley}), we will fix problems (\ref{item:problem-zero}) and (\ref{item:problem-plus}) by representing an $s : \Staged c$ instead with $\fun(s' : \Staged c).\ s \PlusStaged s'$.
This allows us to represent zeros of type $c$ using the identity function that adds nothing and $(\PlusStaged)$ using function composition, both of which need constant time to create.
This will also reduce the injector backpropagators from problem (\ref{item:problem-onehot}) to run in time proportional to the time needed to project their corresponding scalar from the input.
This is not quite constant time yet; instead, it is linear in the maximal nesting depth of the input.

Afterwards, in \cref{sec:mutarrays}, we will fix problem (\ref{item:problem-onehot}) (now fully) by allocating a mutable array for the collection of input scalar cotangents, because this, in a sense, reduces the input nesting depth to a constant 1: the injector backpropagators will be implemented with a single-scalar mutable update.
Furthermore, we will fix problem (\ref{item:problem-map}) by allocating another mutable array for the backpropagators, instead of using a tree map in $\Staged$ that has integers as keys.
Both of these fixes are made possible in a purely functional context by using resource-linear types.

\section{Cayley-transforming the cotangent collector}\label{sec:cayley}

There is a classical trick in functional programming that is sometimes referred to as the ``difference list'' trick, referring to its first popularisation by~\cite{fp-1986-difference-lists} as applied to appending lists.
The idea is that when one has a monoid, that is to say a set (type) $S$ with an identity element $e$ and an associative combination operation $\diamond$, one can represent an element $x \in S$ instead by the partial application (either on the left or on the right, as suits the application) of $\diamond$ to $x$.
Converting from the alternate representation back to the standard one involves just applying the function in hand to the identity element $e$.

\begin{example}
In the classical example, where $S$ is the type of cons-lists with elements of some type $\tau$ (defined inductively as $[] \in S$ and $x \in \tau, l \in S \Rightarrow x :: l \in S$),\footnote{So in Haskell terms, $S = \texttt{[$\tau$]}$.} $e$ is the empty list $[]$ and $\diamond$ is list concatenation ($\listappend$), the aim of the alternate representation is to improve performance when appending many lists in a left-associative order: $((l_1 \listappend l_2) \listappend l_3) \listappend l_4$.
Indeed, with the standard (and only reasonable) definition of list concatenation, given by $[] \listappend l = l$ and $(x :: xs) \listappend l = x :: (xs \listappend l)$, appending $n$ lists of length $m$ in left-associative order takes time $O(n^2m)$.
This is in strong contrast to the right-associative concatenation order ($l_1 \listappend (l_2 \listappend (l_3 \listappend l_4))$), which would take time $O(nm)$, being the correct, and lowest-possible, complexity for cons-lists.

In the alternate ``difference list'' representation, we take for a list $l$ the representation $\fun l'.\ l \listappend l'$ which prepends the list in question to some to-be-determined suffix list.
``Appending'' these representations entails just function composition, and the final, composed result can be converted back to an actual list by applying the resulting function to the empty list.
The insight is that in the final application to the empty list, after evaluating away the function composition closures (in time $O(n)$), the list append operations will always happen in right-associative order, regardless of the order of composition of the difference-list functions: $(((\fun l'.\ l_1 \listappend l') \circ (\fun l'.\ l_2 \listappend l')) \circ (\fun l'.\ l_3 \listappend l'))\ []$ evaluates (in time $O(n)$) to $l_1 \listappend (l_2 \listappend (l_3 \listappend []))$, which then finishes in $O(nm)$.
This means that the whole operation of converting to difference-list representation, composing the functions obtained, and applying the result to the empty list again, always runs in time $O(nm)$, as required.\footnote{Difference lists are used in practice in the \texttt{Show} typeclass in Haskell: \texttt{shows :: Show a => a -> ([Char] -> [Char])}. Indeed, the current standard library (base-4.16.0.0) has \texttt{show x = shows x []} as default implementation.}
\end{example}

In our case, the problematic type $S$ is that of $\Staged$ objects, but the issue is not so much the associativity of their combination---indeed, $(\PlusStaged)$ has many problems but sensitivity to associativity is not one of them---but the fact that we are doing work in the combination operation at all.
However, we can still fruitfully apply the difference list trick, which we will call \emph{Cayley-transforming} $\Staged$, in reference to Cayley's theorem in group theory.\footnote{%
  Cayley's theorem says that for a group $G$, the map that sends $g \in G$ to the permutation induced by left-multiplication with $g$ is an injective group homomorphism into $\mrm{Sym}(G)$, the permutation group over $G$'s underlying set.
  For a monoid $M$, the analogous theorem is that the map that sends $x \in M$ to the function $y \mapsto x \diamond y$ is an injective monoid homomorphism into the monoid of functions from $M$ to $M$.
  The (more recent) Yoneda lemma vastly generalises both variants of Cayley's theorem, but the application to monoids feels closer to the original group case than the more general lemma in category theory.
}
This means that instead of $\Staged c$ we will now work with $\Staged c \ra \Staged c$, and instead of $(\PlusStaged)$ we will now have function composition.

\subsection{Code transformation}

\begin{figure}
  \figheading{On types:}
  \begin{gather*}
    \trans4c[\R] = (\R, (\Int, \R \lra (\Staged c \ra \Staged c))) \qquad
    \textcolor{gray}{\trans4c[()] = ()} \qquad
    \textcolor{gray}{\trans4c[(\sigma,\tau)] = (\trans4c[\sigma], \trans4c[\tau])} \\
    \textcolor{gray}{\trans4c[\sigma \ra \tau] = \trans4c[\sigma] \ra \Int \ra (\trans4c[\tau], \Int)} \qquad
    \textcolor{gray}{\trans4c[\Int] = \Int}
  \end{gather*}
  \figheading{On terms:}
  \begin{align*}
    &\textcolor{gray}{\text{If}\ \Gamma \vdash t : \tau\ \text{then}\ \trans4c[\Gamma] \vdash \trans4c[t] : \Int \ra (\trans4c[\tau], \Int)} \\
    &\text{Same as $\trans3c$, except with `$\id$' in place of $\ZeroStaged$ and `$\circ$' in place of $(\PlusStaged)$.}
  \end{align*}
  \figheading{Updated $\Staged$ interface:}
  \begin{align*}
    &\begin{array}{@{}l@{\ }c@{\ }ll}
      \Staged c &=& (c, \Map\ \Int\ (\R \lra (\Staged c \ra \Staged c), \R)) \\[2mm]
      \StagedRunZero &:& (\Staged c \ra c) \ra c \\
      \StagedRunZero\ f &=& f\ (\underline0, \EmptyMap)
        \qquad\textit{(Note: still $O(|c|)$, but only used once (\cref{fig:wrapper-cayley}))}\\[2mm]
      \StagedCall &:& (\Int, \R \lra (\Staged c \ra \Staged c)) \ra \R \lra (\Staged c \ra \Staged c) \\
      \StagedCall\ (i, f)\ x &=& \fun(c, m).\ (c, \text{if}\ i \not\in m\ \text{then}\ \text{insert}\ \{i \mapsto (f, x)\}\ \text{into}\ m \\
        && \hspace{2.745cm} \text{else} \hspace{0.18cm} \text{update}\ m\ \text{at}\ i\ \text{with}\ (\fun(\mathunderscore, x').\ (f, x + x'))) \\[2mm]
      \StagedMapCot &:& (c \ra c) \lra (\Staged c \ra \Staged c) \\
      \StagedMapCot\ f &=& \fun(c, m).\ (f\ c, m) \\[2mm]
      \ResolveStaged &:& \Staged c \ra c \\
      \ResolveStaged\ (c, m)
        &=&\textcolor{gray}{\mbf{if}\ m\ \text{is empty}} \\
        &&\quad\textcolor{gray}{\mbf{then}\ c} \\
        &&\quad\textcolor{gray}{\mbf{else}\ \mbf{let}\ i = \text{highest key in}\ m} \\
        &&\quad\hphantom{\mbf{else}}\ \textcolor{gray}{\mbf{in}\ \mbf{let}\ (f, a) = \text{lookup}\ i\ \text{in}\ m} \\
        &&\quad\hphantom{\mbf{else}}\ \textcolor{gray}{\mbf{in}\ \mbf{let}\ m' = \text{delete}\ i\ \text{from}\ m} \\
        &&\quad\hphantom{\mbf{else}}\ \textcolor{gray}{\mbf{in}\ \ResolveStaged(}f\ a\ (c, m')\textcolor{gray}{)}
          \qquad\textit{(Look: no $(\PlusStaged)$!)}
    \end{array}
  \end{align*}
  \caption{\label{fig:algo-cayley}
    The Cayley-transformed code transformation, based on \cref{fig:algo-monadic}.
    Grey parts are unchanged.
    Also see \cref{fig:wrapper-cayley} for the corresponding wrapper.
  }
\end{figure}

\begin{figure}
  \figheading{Wrapper:}
  \begin{align*}
    &\begin{array}{@{}l@{\ }c@{\ }l}
      \Interleave4_\tau &:& \forall c.\ (\tau, (\tau \ra \tau) \lra (\Staged c \ra \Staged c)) \ra \Int \ra (\trans4c[\tau], \Int) \\
      \Interleave4_{\R} &=& \fun(x, d).\ \fun i.\ ((x, (i, \lfun(z : \R).\ d\ (\fun(a : \R).\ z + a))) \\
      &&\hspace{1.65cm} , i + 1) \\
      \Interleave4_{()} &=& \fun((), d).\ \fun i.\ ((), i) \\
      \Interleave4_{(\sigma, \tau)} &=& \fun((x, y), d).\ \fun i. \\
      &&\hspace{-0.7cm} \mbf{let}\ (x', i') = \Interleave4_\sigma\ (x, \fun(f : \sigma \ra \sigma).\ d\ (\fun((v, w) : (\sigma, \tau)).\ (f\ v, w)))\ i \\
      &&\hspace{-0.7cm} \mbf{in}\ \mbf{let}\ (y', i'') = \Interleave4_\tau\ (y, \fun(f : \tau \ra \tau).\ d\ (\fun((v, w) : (\sigma, \tau)).\ (v, f\ w)))\ i' \\
      &&\hspace{-0.7cm} \mbf{in}\ ((x', y'), i'') \\
      \Interleave4_{\Int} &=& \fun(n, d).\ \fun i.\ (n, i) \\
      \Interleave4_{\sigma \ra \tau} &=& \text{not defined!}
    \end{array} \\[0.2cm]
    &\begin{array}{@{}l@{\ }c@{\ }l}
      \Deinterleave4_\tau &:& \forall c.\ \trans4c[\tau] \ra (\tau, \tau \lra (\Staged c \ra \Staged c)) \\
      \mathrlap{\text{(Same as $\Deinterleave3$ in \cref{fig:algo-monadic}, except with $\id$ and $(\circ)$ in place of $\ZeroStaged$ and $(\PlusStaged)$)}}
    \end{array} \\[0.2cm]
    &\begin{array}{@{}l}
      \Wrap4 : (\sigma \ra \tau) \rightsquigarrow (\sigma \ra (\tau, \tau \lra \sigma)) \\
      \Wrap4[\lambda(x : \sigma).\ t] = \fun(x : \sigma).\ \mbf{let}\ (x : \trans4\sigma[\sigma], i) = \Interleave4_\sigma\ (x, \StagedMapCot)\ 0 \\
      \hspace{4.412cm} \mbf{in}\ \mbf{let}\ (y, d) = \Deinterleave4_\tau\ (\fst(\trans4\sigma[t]\ i)) \\
      \hspace{4.412cm} \mbf{in}\ (y, \lfun(z : \tau).\ \StagedRunZero\ (\ResolveStaged \circ d\ z))
    \end{array}
  \end{align*}
  \caption{\label{fig:wrapper-cayley}
    The wrapper for the Cayley-transformed code transformation, presented together with \cref{fig:algo-cayley}.
  }
\end{figure}

The updated code transformation is shown in \cref{fig:algo-cayley}, with corresponding wrapper in \cref{fig:wrapper-cayley}.
We can see that the described replacement has indeed been performed: for example, the occurrence of $\Staged c$ in $\trans3c[\R]$ (\cref{fig:algo-monadic}) has been replaced by $\Staged c \ra \Staged c$ in $\trans4c[\R]$.
Note that the notation $\R \lra (\Staged c \ra \Staged c)$ is sensible, because $(\Staged c \ra \Staged c)$ is indeed a monoid (namely under function composition).\footnote{If we must be precise, the $\lra$-arrow was defined as a \emph{commutative} monoid homomorphism, and the full function space $\Staged c \ra \Staged c$ is not a commutative monoid. However, Cayley actually maps us to the submonoid of functions semantically equivalent to $(\fun s'.\ s \PlusStaged s')$ for some $s \in \Staged c$, which \emph{is} a commutative monoid, because $\Staged c$ is.}

In a sense, this whole section describes a well-known and standard trick, and indeed the changes to the inductive code transformation itself ($\trans4c$) amount to just exchanging the $\Staged$ monoid for the endomorphism monoid ($\tau \ra \tau$ under function composition).
However, contrary to the traditional case for difference lists, we are not just Cayley-transforming for the re-association effects, but rather to be able to implement the wrapper and the $\Staged$ interface in a much better way.
Hence, this is what we will focus on in the remainder of this section.


First we will look at the updated wrapper in \cref{fig:wrapper-cayley}; afterwards, we will discuss the updated $\Staged$ interface.
Because $\Interleave4$ now needs to produce $\Staged c$ updaters instead of $\Staged c$ values to juxtapose to scalars, $\Interleave4$ now takes as input not an ``injector'' function of type $\tau \lra \Staged c$ (or, after the Cayley transformation, $\tau \lra (\Staged c \ra \Staged c)$), but instead a function of type $(\tau \ra \tau) \lra (\Staged c \ra \Staged c)$, reminiscent of a Setter optic: given an updater for $\tau$ this argument should build an updater for the whole $\Staged c$ that applies the $\tau$-updater on the current location in $\Staged c$.
This Setter is used in $\Interleave4_\R$, where a backpropagator of type $\R \lra (\Staged c \ra \Staged c)$ is built by using the Setter ($d$, here of type $(\R \ra \R) \lra (\Staged c \ra \Staged c)$) to add the cotangent of this scalar to the correct position of the full $\Staged c$ object.
Note that we give the created backpropagator an integer tag $i$, just like in $\Interleave3$ in \cref{fig:algo-monadic}.
For pairs, $\Interleave4_{(\sigma,\tau)}$ performs just the necessary administration to build up these Setters.

In $\Deinterleave4$, like in the transformation $\trans4c$ itself, we simply replace $\ZeroStaged$ with the identity function and $(\PlusStaged)$ with function composition.
Indeed, the only things we changed since $\Deinterleave1$ from \cref{fig:wrapper-naive} is the types and choice of monoidal zero and plus, and the fact that starting with $\Deinterleave3$ (\cref{fig:algo-monadic}), the new $\Int$ ID of backpropagators is ignored.

Next, in the wrapper $\Wrap4$ itself, we start the interleaving process by passing $\StagedMapCot: (\sigma \ra \sigma) \lra (\Staged \sigma \ra \Staged \sigma)$, which updates the $\sigma$ value contained in the first component of $\Staged \sigma$, as shown in the updated $\Staged$ interface in \cref{fig:algo-cayley}.
Once we have a dualised input and an ID to start off the monadic computation produced by the transformed program, we continue precisely as in $\Wrap3$ in \cref{fig:algo-monadic} until the point where we define the final top-level derivative function, where instead of obtaining a $\Staged \sigma$ from $d(z)$, we get an \emph{updater} for $\Staged \sigma$ to which we can post-compose $\ResolveStaged$ to get a function of type $\Staged \sigma \ra \sigma$, which we apply to a zero initial $\Staged \sigma$ using $\StagedRunZero$ from \cref{fig:algo-cayley}.

Finally, let us briefly analyse the updated $\Staged$ interface in \cref{fig:algo-cayley} that we already used in the text above.
$\StagedRunZero$ still builds a zero of type $c$, but as indicated, this is not a problem for complexity because this zero is only created exactly once (contrary to $\ZeroStaged$ from earlier, which was used many times throughout the computation).
Subsequently, this zero is \emph{added to} by all the $c \ra c$ updaters (lifted using $\StagedMapCot$) created by the injector backpropagators in $\Interleave4$, which are called at input variable references in the source program.
The new version of $\StagedCall$ is precisely $(\PlusStaged)$ from the ``Actual implementation'' in \cref{fig:staged-tree-impl} partially applied to $\StagedCall$ from that same place: we specialised the map union operation to the case where one of the operands is a map containing exactly one key-value pair.
The new primitive $\StagedMapCot$ simply applies the $c$ updater to the first component of the pair that is a $\Staged c$.
Lastly, the definition of $\ResolveStaged$ that we gave in \cref{sec:staged-interface} is changed only in one place: the return value from $f$ is now not a $\Staged c$ object that needs to be added to the running total, but instead an updater function that we can directly apply.

\subsection{Complexity}\label{sec:cayley-complexity}

By Cayley-transforming $\Staged$, we successfully ensured that all the backpropagators created in $\trans4c$ itself perform no more work than the corresponding source terms, and that $\Deinterleave4$ no longer needs to add many $\Staged c$ objects together; in short, we eliminated all problematic uses of $\ZeroStaged$ and $(\PlusStaged)$.
However, as indicated in \cref{sec:staged-complexity}, problems (\ref{item:problem-onehot}) and (\ref{item:problem-map}) with the time complexity still remain: the injector backpropagators (created in $\Interleave4$) still traverse the nesting depth of the input for each individual scalar present in the input, and the $\Map$ operations in $\ResolveStaged$ are still logarithmic.

But notice that the injection backpropagators are essentially functions of type $\R \lra (c \ra c)$; these are lifted to $\R \lra (\Staged c \ra \Staged c)$ using $\StagedMapCot$ which just applies a $c \ra c$ function to the first component of the pair in a $\Staged c$.
The point of these $\R \lra (c \ra c)$ functions is to add the $\R$ argument to a particular scalar contained in $c$, and this is accomplished by traversing the nested tuples in $c$ leading to that particular scalar.
Clearly, it would be better if it would take only constant time to add a single scalar to one position in the input cotangent that we are collecting.
Such constant-time random access is only possible with a mutable array, and to use mutable arrays in a purely functional setting, we need a resource-linear typing of the functions that perform these updates to the array.
Fortunately, as we will see in \cref{sec:mutarrays}, we can give the $\Staged c \ra \Staged c$ updater functions a resource-linear type, which allows us to use a mutable array for the input cotangent collection.

The second remaining issue is that $\ResolveStaged$ performs some logarithmic $\Map$ operations; however, the $\Map$ in question has type `$\Map\ \Int\ (\R \lra (\Staged c \ra \Staged c))$' and is thus keyed by integers---and in fact, these integers are even consecutive and starting from 0, because they are precisely the ID values we generate for the backpropagators in transformed code.
In imperative languages, a $\Map$ from consecutive integer values is more naturally represented as an array, and like above, this is precisely what we shall do.

\section{Utilising resource-linear types to shave off log factors}\label{sec:mutarrays}

By introducing resource-linearity into our type system, we now have three flavours of function arrows: regular functions ($\ra$), monoid-linear functions ($\lra$) and the new resource-linear functions, which we write as $\rlra$.
Where regular functions used $\fun$ and monoid-linear functions used $\lfun$, resource-linear functions use $\rlfun$.
Both monoid-linearity and resource-linearity are essential for achieving the right time complexity in this paper, but in somewhat different ways: monoid-linearity shows that the optimisations that we perform preserve semantics, while resource-linearity shows that the transition to mutable arrays preserves referential transparency.
In Haskell, resource-linearity must be explicitly annotated in the output program of the code transformation for GHC to accept the code, while monoid-linearity need not be reflected in the implementation.
In an hypothetical implementation in OCaml, where mutable updates are allowed anywhere, neither form of linearity would need to be reflected.

The type system of resource-linearity that we use is that of Linear Haskell~\cite{fp-2018-linear-haskell}, and is similar in intent, though not identical in design, to that of the Rust language.\footnote{In Rust, values may be implicitly dropped, making its version of linear types really \emph{affine} instead of purely linear. Furthermore, Rust extends the system with a complex but convenient system of ownership and lifetime tracking.}
For an introduction to Linear Haskell, we refer to the cited article.

For working with mutable arrays using resource-linearity, we use a standard interface taken from the \texttt{linear-base}\footnote{\url{https://hackage.haskell.org/package/linear-base-0.2.0/docs/Data-Array-Mutable-Linear.html}} Haskell library.
(For background information on implementation as well as usage idioms, we refer again to~\cite{fp-2018-linear-haskell}.)
This library is written by the authors of Linear Haskell as a general standard library for usage of linear types in Haskell.
The subset of the array interface that we use is listed and briefly explained in \cref{fig:array-interface}.\footnote{In \texttt{linear-base}, the final $\ra \sigma$ in the type of $\arralloc$ is instead $\rlrafootnote \ur \sigma$, but this is relevant only if the $\Array \tau \rlrafootnote \ur \sigma$ function closes over a linear value, which ours does not. Hence we choose the simpler presentation.}
In the same figure, we also give a derived operation $\arrmodify$.

\begin{figure}
  \figheading{Array interface:}
  \begin{gather*}
    \begin{array}{@{}l@{\;}c@{\;}l@{}}
      \arralloc &:& (\mit{length} : \Int) \ra (\mit{initval} : \tau) \ra (\Array \tau \rlra \ur \sigma) \ra \sigma \\
        \mathrlap{\quad \text{Allocate an array of the given $\mit{length}$ filled with $\mit{initval}$ in all positions, and pass it to the}} \\
        \mathrlap{\quad \text{given function. Referential transparency is ensured because linearity guarantees that the}} \\
        \mathrlap{\quad \text{$\Array \tau$ cannot be returned as part of the $\ur \sigma$.}} \\
      \arrallocBeside &:& (\mit{length} : \Int) \ra (\mit{initval} : \tau) \ra (witness : \Array \sigma) \rlra (\Array \sigma, \Array \tau) \hspace{0.8cm}{} \\
        \mathrlap{\quad \text{Using a ``linearity witness'', allocate another array. In particular: $\arrallocBeside\ l\ x\ w$ returns}} \\
        \mathrlap{\quad \text{two arrays: $w$ and a newly allocated array of length $l$ filled with $x$es.}} \\
      \arrget &:& (\mit{index} : \Int) \ra \Array \tau \rlra (\ur \tau, \Array \tau) \\
        \mathrlap{\quad \text{Note: The returned value of type $\tau$ does not need to be used linearly. While this means that}} \\
        \mathrlap{\quad \text{this interface is unsuitable for arrays of (mutable) arrays, we will not need those here.}} \\
      \arrset &:& (\mit{index} : \Int) \ra (\mit{value} : \tau) \ra \Array \tau \rlra \Array \tau \\
      \arrsize &:& \Array \tau \rlra (\ur \Int, \Array \tau) \\
      \arrdealloc &:& \Array \tau \rlra \Array \sigma \rlra \Array \sigma \\
        \mathrlap{\quad \text{$\arrdealloc\ a\ b$ deallocates $a$ and returns $b$. (This is \texttt{lseq} from \texttt{linear-base} with a more}} \\
        \mathrlap{\quad \text{specific type.)}} \\
      \arrfreeze &:& \Array \tau \rlra \ur (\IArray \tau) \\
        \mathrlap{\quad \text{Permanently convert a mutable array to an immutable array. $\IArray$ corresponds to}} \\
        \mathrlap{\quad \text{\texttt{Vector} in the Haskell \texttt{linear-base} API.}} \\
      (\iarridx) &:& \IArray \tau \ra \Int \ra \tau \\
        \mathrlap{\quad \text{Index an immutable array---no linearity required. (\texttt{(Data.Vector.!)} in Haskell.)}}
    \end{array}
  \end{gather*}
  \figheading{Sequencing:}
  \begin{gather*}
    \begin{array}{l}
      \rllet\ (x_1, \ldots, x_n) = t_1\ \rlin\ t_2 \\
        \quad \text{Compute $t_1$, bind (and, if $n > 1$, destructure) its result to $x_1,\ldots,x_n$, and finally compute $t_2$.} \\
        \quad \text{$t_1$ is allowed to consume linearly bound values, as long as they are unused in $t_2$. In Haskell} \\
        \quad \text{using GHC 9.2, this is written as \texttt{case $t_1$ of ($x_1$,$\ldots$,$x_n$) -> $t_2$}.} \\
    \end{array}
  \end{gather*}
  \figheading{Derived array operations}
  \begin{gather*}
    \begin{array}{@{}l@{\;}c@{\;}l@{}}
      \arrmodify &:& (\mit{index} : \Int) \ra (\tau \ra \tau) \ra \Array \tau \rlra \Array \tau \\
      \arrmodify\ i\ f &=& \rlfun(a : \Array \tau).\ \rllet\ (\ur x, a') = \arrget\ i\ a\ \rlin\ \arrset\ i\ (f\ x)\ a' \\
      \mathrlap{\quad \text{Modify the array at the given index.}}
    \end{array}
  \end{gather*}
  \caption{\label{fig:array-interface}
    Features from Linear Haskell that we use, including a subset of \texttt{Data.Array.Mutable.Linear} from the Haskell library \texttt{linear-base}, as well as a linear let-binding.
    We write the Haskell type \texttt{Ur} (``of course'' or ``unrestricted'') as $\ur$, in types as well as in pattern matches.
  }
\end{figure}

\paragraph{Transformation using arrays}
As introduced at in \cref{sec:cayley-complexity}, we will use arrays for two things: as replacement for the $c$ value in a $\Staged c$ (in which the final cotangent is collected by invocation of the injector backpropagators), and as a replacement for the staging $\Map$ in a $\Staged c$.
The array replacing $c$ is of type $\Array \R$, indicating a variably-sized array of scalars; we are allowed to make this replacement because we are, indeed, only interested in the scalars in $c$.\footnote{The non-scalar information in the input is contained in the $\IArray \R \ra \sigma$ output from $\Interleave5_\sigma$; see \cref{fig:algo-mutarrays} and the main text.}
%
Hence, we change $\Staged c$ from:
\[ (c\hspace{0.98cm}, \Map\ \Int\ (\R \lra (\Staged c \ra \Staged c), \R)) \]
to:
\[ (\Array \R, \Array{} \hspace{0.3cm}(\R \lra (\Staged c \rlra \Staged c), \R)) \]
As previously stated, we also introduce a resource-linear arrow in the updater functions to allow mutation of the arrays in those updater functions.
Now, because this type no longer structurally depends on the choice of $c$, we rename $\Staged c$ to $\State$:
\[ \State = (\Array \R, \Array{} \hspace{0.3cm}(\R \lra (\State\hspace{0.45cm} \rlra \State\hspace{0.45cm}), \R))\hspace{1.12cm} \]
This loss of dependence on the choice of $c$ is because that information is now (only) contained in the \emph{length} of the cotangent collection array (of type $\Array \R$), which is not reflected on the type level using our array interface.

\subsection{Code transformation}

\begin{figure}
  \figheading{On types:}
  \begin{gather*}
    \trans5{}[\R] = (\R, (\Int, \R \lra (\State \rlra \State))) \qquad
    \textcolor{gray}{\trans5{}[()] = ()} \qquad
    \textcolor{gray}{\trans5{}[(\sigma,\tau)] = (\trans5{}[\sigma], \trans5c[\tau])} \\
    \textcolor{gray}{\trans5{}[\sigma \ra \tau] = \trans5{}[\sigma] \ra \Int \ra (\trans5{}[\tau], \Int)} \qquad
    \textcolor{gray}{\trans5{}[\Int] = \Int}
  \end{gather*}
  \figheading{On terms:}
  \begin{align*}
    &\textcolor{gray}{\text{If}\ \Gamma \vdash t : \tau\ \text{then}\ \trans5{}[\Gamma] \vdash \trans5{}[t] : \Int \ra (\trans5{}[\tau], \Int)} \\
    &\text{Same as $\trans4c$. (`$\id$' now has a resource-linear type, $\State \rlra \State$. Note further that the usages} \\
    &\text{of $\StagedCall$ remain type-correct.)}
  \end{align*}
  \figheading{New $\State$ interface, replacing $\Staged$:}
  \begin{align*}
    &\begin{array}{@{}l@{\ }c@{\ }ll}
      \State &=& (\Array \R, \Array{} (\R \lra (\State \rlra \State), \R)) \\[2mm]
      \StateAlloc &:& \Int \ra \Int \ra (\State \rlra \ur c) \ra c \\
      \StateAlloc\ i_{\text{inp}}\ i_{\text{out}}\ f &=& \texttt{alloc}\ i_{\text{inp}}\ 0\ (f \circ \texttt{allocBeside}\ i_{\text{out}}\ (\lfun(z : \R).\ \id, 0)) \\[2mm]
      \StagedCall &:& (\Int, \R \lra (\State \rlra \State)) \ra \R \lra (\State \rlra \State) \\
      \StagedCall\ (i, f)\ a &=& \rlfun(c, m).\ (c, \arrmodify\ i\ (\fun (\mathunderscore, a').\ (f, a + a'))\ m) \\[2mm]
      \InputCot &:& \Int \ra \R \lra (\State \rlra \State) \\
      \InputCot\ i\ a &=& \rlfun(c, m).\ (\arrmodify\ i\ (\fun(a' : \R).\ a + a')\ c, m) \\[2mm]
      \ResolveState &:& \Int \ra \State \rlra \Array \R \\
      \ResolveState\ i_{\text{out}}
        &=& \rlfun(c, m).\ \rllet\ (c, m) = \mit{loop}\ (i_{\text{out}} - 1)\ (c, m) 
        \ \rlin\ \arrdealloc\ m\ c \\
      \quad\text{where}\ \mit{loop}\ &:& \Int \ra \State \rlra \State \\
      \hphantom{\quad\text{where}}\ \mit{loop}\ 0 &=& \rlfun(s : \State).\ s \\
      \hphantom{\quad\text{where}}\ \mit{loop}\ i &=& \rlfun(c, m).\ \rllet\ (\ur (f, a), m) = \texttt{get}\ i\ m 
        \ \rlin\ \mit{loop}\ (i - 1)\ (f\ a\ (c, m))
    \end{array}
  \end{align*}
  \figheading{Wrapper:}
  \begin{align*}
    &\begin{array}{@{}l@{\;}c@{\;}l@{}}
      \Interleave5_\tau &:& \tau \ra \Int \ra ((\trans5{}[\tau], \IArray \R \ra \tau), \Int) \\
      \Interleave5_\R &=& \fun x.\ \fun i.\ (((x, (i, \InputCot\ i)), \fun a.\ a \iarridx i), i + 1) \\
      \Interleave5_{()} &=& \fun ().\ \fun i.\ (((), \fun a.\ ()), i) \\
      \Interleave5_{(\sigma, \tau)} &=& \fun (x, y).\ \fun i.\ \mbf{let}\ ((x', f_1), i') = \Interleave5_\sigma\ x\ i \\
      &&\hspace{1.6cm} \mbf{in}\ \mbf{let}\ ((y', f_2), i'') = \Interleave5_\tau\ y\ i' \\
      &&\hspace{1.6cm} \mbf{in}\ (((x', y'), \fun a.\ (f_1\ a, f_2\ a)), i'') \\
      \Interleave5_\Int &=& \fun n.\ \fun i.\ ((n, \fun a.\ n), i) \\
      \Interleave5_{\sigma \ra \tau} &=& \text{not defined!}
    \end{array} \\[0.2cm]
    &\begin{array}{@{}l}
      \Deinterleave5_\tau : \trans5{}[\tau] \ra (\tau, \tau \lra (\State \rlra \State)) \\
      \text{Same as $\Deinterleave3$ except for types.}
    \end{array} \\[0.2cm]
    &\begin{array}{@{}l}
      \Wrap5 : (\sigma \ra \tau) \rightsquigarrow (\sigma \ra (\tau, \tau \lra \sigma)) \\
      \Wrap5[\lambda(x : \sigma).\ t] = \fun(x : \sigma). \\
      \qquad \mbf{let}\ ((x : \trans5{}[\sigma], \mit{rebuild} : \IArray \R \ra \sigma), i) = \Interleave5_\sigma\ x\ 1 \\
      \qquad \mbf{in}\ \mbf{let}\ (y', i') = \trans5{}[t]\ i \\
      \qquad \mbf{in}\ \mbf{let}\ (y, d : \tau \lra (\State \rlra \State)) = \Deinterleave5_\tau\ y' \\
      \qquad \mbf{in}\ (y, \lfun(z : \tau).\ \mit{rebuild}\ (\StateAlloc\ i\ i'\ (\arrfreeze \circ \ResolveState\ i' \circ d\ z)))
    \end{array}
  \end{align*}
  \caption{\label{fig:algo-mutarrays}
    Code transformation plus wrapper using mutable arrays, modified from \cref{fig:algo-cayley,fig:wrapper-cayley}.
    Grey parts are unchanged.
  }
\end{figure}

Using this new type, we change the code transformation once more, this time from \cref{fig:algo-cayley} to the version given in \cref{fig:algo-mutarrays}.
While the transformation on types and terms simply gains resource-linearity on the arrows in scalar backpropagators, and thus does not materially change, some important changes occur in the $\State$ interface (previously the $\Staged c$ interface) and the wrapper.
Let us first look at the algorithm from the top, by starting with $\Wrap5$; after understanding the high-level idea, we explain how the other components work.

In basis, $\Wrap5$ does the same as $\Wrap4$ from \cref{fig:wrapper-cayley}: interleave injector backpropagators with the input of type $\sigma$, execute the transformed function body using the interleaved input, and then deinterleave the result.
However, because we now represent the final cotangent not directly as a value of type $\sigma$ in a $\Staged \sigma$ but instead as an array of only the embedded scalars ($\Array \R$), some more work needs to be done.

Firstly, $\Interleave5_\sigma$ (monadically) produces, in addition to the interleaved input, also a \emph{rebuilder} of type $\IArray \R \ra \sigma$.
This rebuilder takes an array with precisely as many scalars as were in the input, and produces a value of type $\sigma$ with the structure (and discrete-typed values) of the input, but the scalars from the array.
The mapping between locations in $\sigma$ and indices in the array is the same as the numbering performed by $\Interleave5$.

Having $x$, $\mit{rebuild}$ and $i$ (the next available ID), we execute the transformed term $t$ monadically (with $x$ in scope), resulting in an output $y' : \trans5{}[\tau]$.
This output we deinterleave to $y : \tau$ and $d : \tau \lra (\State \rlra \State)$.

The final result then consists of the regular function result ($y$) as well as the top-level derivative function of type $\tau \lra \sigma$.
This $\tau$ we can pass to $d$ to get a $\State$ updater that (because of how $\Deinterleave5$ works) stages calls to the top-level backpropagators contained in $y'$ in a $\State$.
Assuming that we can pass $d\ z$ an empty $\State$, we then use $\ResolveState$ (corresponding to $\ResolveStaged$ from \cref{fig:algo-cayley}) to propagate the cotangent contributions backwards, by invoking each backpropagator in turn in descending order of IDs.
Like before in the Cayley version, those backpropagators update the state (now mutably) to record their own contributions to (i.e.\ invocations of) other backpropagators.
As listed in \cref{fig:algo-mutarrays}, $\ResolveState$ does not return the entire state but only the cotangent collection array of type $\Array \R$ (corresponding to the $c$ value in a $\Staged c$ for the Cayley version); the other array is deallocated before returning.
Passing this $\Array \R$ to $\arrfreeze$ gives us an unrestricted $\IArray \R$ containing the cotangents of all scalars in the top-level input.

At this point we have built $z : \tau \vdash (\arrfreeze \circ \ResolveState \circ d\ z) : \State \rlra \ur (\IArray \R)$, which we need to pass an empty $\State$.
This we do using $\StateAlloc$ from \cref{fig:algo-mutarrays}, which uses an idiom in Linear Haskell (also used in $\arralloc$): to allow code to work with a mutable data structure, enforce that said code takes the mutable data structure resource-linearly to an \emph{unrestricted} return value; with that typing, the mutable data structure cannot escape through the return value, meaning that the mutation is invisible from outside, as required.
The additional $i$ and $i'$ arguments to $\StateAlloc$ are the sizes of the arrays to allocate: the cotangent array of type $\Array \R$ will be indexed by the IDs of the scalars in the interleaved input (which are in $\{0,\ldots,i-1\}$), and the backpropagator staging array (of type $\Array{} (\R \lra (\State \rlra \State), \R)$) will be indexed by the IDs of all backpropagators created in both $\Interleave5$ and $\trans5{}[t]$ (which are in $\{0,\ldots,i'-1\}$).
Hence, $i$ and $i'$ are suitable array sizes.

Finally, the $\IArray \R$ returned by $\arrfreeze$ through $\StateAlloc$ is passed to $\mit{rebuild}$ from $\Interleave5$ to put all scalar cotangents in the correct locations in the input; the resulting final cotangent is returned.

\paragraph{Implementation of the components}
Having discussed the high-level sequence of operations, let us briefly discuss the implementation of the $\State$ interface and (de)interleaving.
In $\Interleave5$, instead of passing structure information down in the form of a setter ($(\tau \ra \tau) \lra (\Staged c \ra \Staged c)$) like we did in $\Interleave4$ in \cref{fig:wrapper-cayley}, we build structure information up in the form of a getter ($\IArray \R \ra \tau$).
This results in a somewhat more compact presentation, but in some sense the same information is still communicated.

The program text of $\Deinterleave5$ is again unchanged, because it is agnostic about the codomain of the backpropagators (apart from being a monoid, in this case over $\id$ and $(\circ)$).

On the $\Staged$ interface the transition to mutable arrays had a significant effect.
The role of $\StagedRunZero$ (\cref{fig:algo-cayley}) is now fulfilled by $\StateAlloc$, which uses $\arralloc$ to allocate the cotangent collection array of size $i_{\text{inp}}$ filled with zeros, and then $\arrallocBeside$ to allocate the backpropagator staging array of size $i_{\text{out}}$ filled with zero-backpropagators with an accumulated argument of zero.
(Recall \cref{fig:array-interface} for the types of $\arralloc$ and $\arrallocBeside$.)
Note that the work that was performed by the $\underline0$ of type $c$ in $\StagedRunZero$ is now performed by $\arralloc$.

$\StagedCall$ has essentially the same type, but its implementation differs: instead of performing a logarithmic-complexity immutable $\Map$ update, we perform a constant-time mutable update on the backpropagator staging array.
Note that, unlike in \cref{fig:algo-cayley}, there is no special case if $i$ is not yet in the array, because unused positions are already filled with zeros.

$\InputCot$ takes the place of $\StagedMapCot$, except we have specialised $\InputCot$ with the knowledge that all relevant $c \ra c$ functions add a particular scalar to a particular index in the input, and that these functions can hence be defunctionalised to a pair $(\Int, \R)$.
The monoid-linearity here is in the real scalar, as it was before, hence the placement of the $\lra$-arrow.

Finally, $\ResolveState$ takes the place of $\ResolveStaged$; it takes an additional $\Int$ argument that gives the output ID of $\trans5{}[t]$, i.e.\ one more than the largest ID generated.
$\mit{loop}$ then performs the loop that $\ResolveStaged$ did, iterating over all IDs in descending order and applying the state updaters in the backpropagator staging array one-by-one to the state.
After the loop is complete, the backpropagator staging array is deallocated and the cotangent collection array is returned, to be passed to $\arrfreeze$ in $\Wrap5$.

\subsection{Complexity}

Let $I$ denote the size of the input and $T$ the runtime of the original program.
We can observe the following:
\begin{itemize}
\item
  The number of operations performed by $\trans5{}[t]$ (which is the same as $\trans3c[t]$ from \cref{fig:algo-monadic} except with $\ZeroStaged$ and $(\PlusStaged)$ replaced with $\id$ and $(\circ)$, respectively) is only a constant factor times the number of operations performed by $t$, and hence in $O(T)$.
  This was already observed in \cref{sec:staged-complexity}.

\item
  The number of backpropagators created in the course of executing $\trans5{}[t]$ is also in $O(T)$.
  This is clear.

\item
  The number of operations performed in any one backpropagator is constant.
  This is new, and only true because $\id$ (replacing $\ZeroStaged$), $(\circ)$ (replacing $(\PlusStaged)$), $\InputCot$ (replacing $\StagedMapCot$ on an injector function) and $\StagedCall$ from \cref{fig:algo-mutarrays} are all constant-time.

\item
  Hence, because every backpropagator is invoked at most once, the amount of work performed by calling the top-level input backpropagator is again in $O(T)$.

\item
  Finally, the (non-constant-time) extra work performed in $\Wrap5$ is interleaving ($O(I)$), deinterleaving ($O(\text{size of output})$ and hence $O(T)$), resolving ($O(T)$) and rebuilding ($O(I)$); all this work is in $O(T + I)$.
\end{itemize}
Hence, calling $\Wrap5[t]$ with an argument and calling its returned top-level derivative once takes time $O(T + I)$, i.e.\ at most proportional to the runtime of calling $t$ with the same argument, plus the size of the argument itself.
As discussed in \cref{sec:rev-ad-complexity}, this is the correct time complexity for an efficient reverse AD algorithm.

\section{Final improvements to the algorithm}\label{sec:improve}

Despite having achieved the correct time complexity, the code transformation in \cref{fig:algo-mutarrays} can still be made more frugal with resources as described in \cref{sec:improve-one-array,sec:improve-defunctionalisation}.
Afterwards, in \cref{sec:improve-taping}, we show how the algorithm relates to standard tape-based reverse AD, and in \cref{sec:relation-krawiec} how it relates to \cite{ad-2021-krawiec-kmett-ad}.

\subsection{Dropping the cotangent collection array}\label{sec:improve-one-array}

Recall that the final transformation $\trans5{}$ used two mutable arrays threaded through the backpropagators in the $\State$ pair: a cotangent collector array of type $\Array \R$ and a backpropagator call staging array of type $\Array{} (\R \lra (\State \rlra \State), \R)$.
The first array is modified by $\InputCot$, and the second by $\StagedCall$.
No other functions modify these arrays.

Looking at the uses of $\InputCot$ in the algorithm, we see that it occurs only in $\Interleave5_\R$ as the backpropagator paired up with scalars in the input of the program.
These backpropagators have ID $i$ and add their received argument to index $i$ in the cotangent collection array.
This means that if $(c, m)$ is the return value of the call to $\mit{loop}$ in $\ResolveState$, we have $c[i] = \snd(m[i])$ for all $i$ for which $c[i]$ is defined.
Therefore, the cotangent collection array is actually unnecessary: its information is directly readable from the backpropagator staging array.

With this knowledge, we can change \cref{fig:algo-mutarrays} to set $\State = \Array{} (\R \lra (\State \rlra \State), \R)$ and to set $\InputCot\ i\ a = \id : \State \rlra \State$; the other primitives change as necessary.
$\ResolveState$ will now not call $\arrdealloc$ any more and simply return $m$ directly; the rebuild functions returned by $\Interleave5_\R$ take the second projection of the array element before returning it.

\subsection{Defunctionalisation of backpropagators}\label{sec:improve-defunctionalisation}

In the core code transformation ($\trans5{}$, excluding the wrapper), all backpropagators are of type $\R \lra (\State \rlra \State)$, and, like observed earlier in \cref{sec:staging}, these backpropagators come in only a limited number of forms:
\begin{enumerate}
\item\label{item:backprop-form-inputcot}
  $(\lfun(z : \R).\ \InputCot\ i\ z)$, as created in $\Interleave5_\R$, reduced to $(\lfun(z : \R).\ \id)$ in \cref{sec:improve-one-array};
\item\label{item:backprop-form-scalar}
  $(\lfun(z : \R).\ \id)$, as created in $\trans5{}[r]$ for scalar constants;
\item\label{item:backprop-form-operator}
  $(\lfun(z : \R).\ \StagedCall\ d_1\ (\partial_1 \mit{op}(x_1,\ldots,x_n)(z)) \circ \cdots \circ \StagedCall\ d_n\ (\partial_n \mit{op}(x_1,\ldots,x_n)(z)))$, as created in $\trans5{}[\mit{op}(x_1,\ldots,x_n)]$.
\end{enumerate}
Furthermore, the information contained in an operator backpropagator of form (3) can actually be described without reference to the value of its argument $z$: because our operators return a single scalar (as opposed to e.g.\ a vector), we have $\frac{\partial f(\mit{op}(x_1,\ldots,x_n))}{\partial x_i} = \frac{\partial f(u)}{\partial u} \cdot \frac{\mit{op}(x_1,\ldots,x_n)}{\partial x_i}$, which can also be written as $\partial_i \mit{op}(x_1,\ldots,x_n)(z) = z \cdot \partial_i \mit{op}(x_1,\ldots,x_n)(1)$.

Hence, we can defunctionalise~\cite{fp-1998-defunctionalisation} and change all occurrences of the type $\R \lra (\State \rlra \State)$ to $\Contrib$, where $\Contrib = [(\Int, \Contrib, \R)]$: a list of triples of an integer, a recursive $\Contrib$ structure, and a scalar.
(The recursive $\Contrib$ structures are assumed to have sharing, similarly to how the references to existing backpropagators in the closures of operator backpropagators (\ref{item:backprop-form-operator}) already had sharing.)
The meaning of $[(i_1, \mit{cb}_1, a_1), \ldots, (i_n, \mit{cb}_n, a_n)]$ of type $\Contrib$ would then be:
\begin{equation*}
\lfun(z : \R).\ \StagedCall\ (i_1, \mit{cb}_1)\ (z \cdot a_1) \circ \cdots \circ \StagedCall\ (i_n, \mit{cb}_n)\ (z \cdot a_n)
\end{equation*}

If we perform the type replacement and the defunctionalisation, $\InputCot$ and $\StagedCall$ disappear, backpropagators of forms (\ref{item:backprop-form-inputcot}) and (\ref{item:backprop-form-scalar}) become $[]$ (the empty list) and those of form (\ref{item:backprop-form-operator}) become:
\begin{equation*}
[(\fst\ d_1, \snd\ d_1, \partial_1 \mit{op}(x_1,\ldots,x_n)(1)), \ldots, (\fst\ d_n, \snd\ d_n, \partial_n \mit{op}(x_1,\ldots,x_n)(1))]
\end{equation*}
$\ResolveState$ then interprets a list of such $(i, \mit{cb}, a)$ by iterating over the list and for each such triple, replacing $(\mit{cb}', a')$ at index $i$ in the staging array with $(\mit{cb}, a' + a)$.

With this defunctionalisation step, we have removed a significant source of newly-introduced lambda abstractions, and hence made the higher-orderness structure much more similar to the original program.
This might ease application of the reverse AD algorithm to languages that lack full support for higher-order functions.

The performance impact of this defunctionalisation step is dependent on the implementation, but will be limited to a constant factor.

\subsection{It was taping all along?}\label{sec:improve-taping}


As a final note, let us look at the final algorithm from a slightly broader perspective.
After the improvements from \cref{sec:improve-one-array,sec:improve-defunctionalisation}, what previously was a tree of (staged) calls to backpropagator functions is now a tree of $\Contrib$ values with attached IDs\footnote{Note that we now have $\trans5{}[\R] = (\R, (\Int, \Contrib))$, the integer being the ID of the $\Contrib$ value.} that are interpreted by $\ResolveState$.
This interpretation (eventually) writes the $\Contrib$ value with ID $i$ to index $i$ in the staging array (possibly multiple times), and furthermore accumulates argument cotangents in the second component of the pairs in the staging array.
While the argument cotangents must be accumulated in reverse order of program execution (indeed, that is the whole point of \emph{reverse} AD), the mapping from ID to $\Contrib$ value can be fully known in the forward pass: the partial derivatives of operators, $\partial_i \mit{op}(x_1,\ldots,x_n)(1)$, can be computed in the forward pass already.

This means that if we change the ID generation monad that the differentiated code already lives in (which is a state monad with a single $\Int$ as state) to additionally carry the staging array, and furthermore change the monad to thread its state through resource-linearly,\footnote{This is possible and results in a linear variant of standard Haskell monads, as described in~\cite{fp-2018-linear-haskell}.} we can already compute the $\Contrib$ lists and write them to the array in the forward pass.
All that $\ResolveState$ then has to do is loop over the array in reverse order (as it already does) and add cotangent contributions to the correct positions in the array according to the $\Contrib$ lists that it finds there.

At this point, there is no meaningful difference any more between this algorithm and what is classically known as taping: we have a tape (the staging array) that we write the performed operations to in the forward pass (automatically growing the array as necessary)---although the tape entries are the already-differentiated operations in this case, and not the original ones.
In this way, we have related the naive version of dual-numbers reverse AD, which admits neat correctness proofs, to the classical, very imperative approach to reverse AD based on taping, which is used in industry-standard implementations of reverse AD (e.g. PyTorch~\cite{ad-2017-pytorch}).\footnote{%
  In terms of performance, the difference between this full taping approach and the algorithm in \cref{fig:algo-mutarrays} is limited to a constant factor.
  Which is actually faster in practice may depend on various language, compiler or platform properties.
}

\subsection[{Relation to [Krawiec et al.\ 2022]}]{Relation to~\cite{ad-2021-krawiec-kmett-ad}}\label{sec:relation-krawiec}

\begin{figure}
  \begin{center}
  \begin{subfigure}[b]{0.48\textwidth}
    \begin{tikzpicture}
      \node at (0, 0) {$\textcolor{gray}{(x, (1,\,} \lfun d.\ ... \textcolor{gray}{))}$};
      \node at (2.5, 0) {$\textcolor{gray}{(y, (2,\,} \lfun d.\ ... \textcolor{gray}{))}$};
      \node at (2.5, -1) {$\textcolor{gray}{(z, (3,\,} \lfun d.\ \Box\ d \circ \Box\ d\textcolor{gray}{))}$};
      \node at (1, -2) {$(x\cdot z, (4, \lfun d.\ \Box\ (z\cdot d) \circ \Box\ (x \cdot d)))$};
      \draw[->] (2.531, -0.99) .. controls (2.4, -0.4) and (0.3, -0.9) .. (0.1, -0.2);
      \draw[->] (3.351, -0.99) .. controls (3.2, -0.7) and (2.9, -0.4) .. (2.6, -0.2);
      \draw[->] (2.045, -2) .. controls (2.16, -1.8) and (2.17, -1.4) .. (2.07, -1.2);
      \draw[->] (0.54, -2) .. controls (0.51, -1.64) and (0.193, -0.56) .. (0.1, -0.2);
    \end{tikzpicture}
    \caption{\label{subfig:sharing-structure-before}
      Before defunctionalisation
    }
  \end{subfigure}
  \hfill
  \begin{subfigure}[b]{0.48\textwidth}
    \begin{tikzpicture}
      \node at (-0.2, 0) {$\textcolor{gray}{(x, (1,\,} []\textcolor{gray}{))}$};
      \node at (2.43, 0) {$\textcolor{gray}{(y, (2,\,} []\textcolor{gray}{))}$};
      \node at (2.8, -1) {$\textcolor{gray}{(z, (3,\,} [(1, \Box, 1.0), (2, \Box, 1.0)]\textcolor{gray}{))}$};
      \node at (1, -2) {$(x\cdot z, (4, [(1, \Box, z), (3, \Box, x)]))$};
      \draw[->] (2.24, -0.99) .. controls (2.12, -0.4) and (0.3, -0.9) .. (0.1, -0.2);
      \draw[->] (3.747, -0.99) .. controls (3.546, -0.7) and (3.15, -0.4) .. (2.85, -0.2);
      \draw[->] (2.15, -2) .. controls (1.81, -1.8) and (1.604, -1.4) .. (1.624, -1.2);
      \draw[->] (0.93, -2) .. controls (0.88, -1.64) and (0.193, -0.56) .. (0.1, -0.2);
    \end{tikzpicture}
    \caption{\label{subfig:sharing-structure-after}
      After defunctionalisation
    }
  \end{subfigure}
  \end{center}
  \caption{\label{fig:sharing-structure}
    The sharing structure before and after defunctionalisation.
    $\StagedCall$ is elided here; in \cref{subfig:sharing-structure-before}, the backpropagator calls are depicted as if they are still normal calls.
    Boxes ($\Box$) are the same in-memory value as the value their arrow points to; two boxes pointing to the same value indicates that this value is \emph{shared}: referenced in two places.
  }
\end{figure}
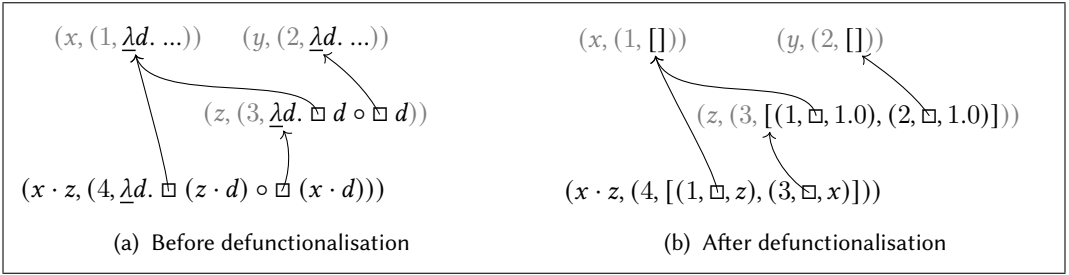

Suppose we differentiate the following simple program:
\begin{align*}
  \fun(x, y).\ \mbf{let}\ z = x + y\ \mbf{in}\ x \cdot z
\end{align*}
using the final algorithm of \cref{sec:mutarrays}, given in \cref{fig:algo-mutarrays}.
The return value from the $\trans5{}$-transformed code (when applied to the output from $\Interleave5$) has the sharing structure shown in \cref{subfig:sharing-structure-before}.
This shows how the backpropagators refer to each other in their closures.

\begin{figure}
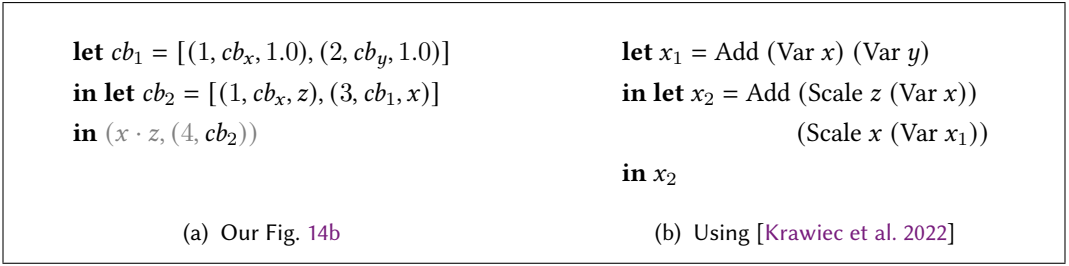

  \begin{center}
  \begin{subfigure}[b]{0.48\textwidth}
    \begin{align*}
      &\mbf{let}\ \mit{cb}_1 = [(1, \mit{cb}_x, 1.0), (2, \mit{cb}_y, 1.0)] \\
      &\mbf{in}\ \mbf{let}\ \mit{cb}_2 = [(1, \mit{cb}_x, z), (3, \mit{cb}_1, x)] \\
      &\mbf{in}\ \textcolor{gray}{(x \cdot z, (4,\,} \mit{cb}_2 \textcolor{gray}{))} \\
      &~
    \end{align*}
    \caption{\label{subfig:tape-program-dualrev}
      Our \cref{subfig:sharing-structure-after}
    }
  \end{subfigure}
  \hfill
  \begin{subfigure}[b]{0.48\textwidth}
    \begin{align*}
      &\mbf{let}\ x_1 = \mrm{Add}\ (\mrm{Var}\ x)\ (\mrm{Var}\ y) \\
      &\mbf{in}\ \mbf{let}\ x_2 = \mrm{Add}\ (\mrm{Scale}\ z\ (\mrm{Var}\ x)) \\
      &\hspace{2.315cm} (\mrm{Scale}\ x\ (\mrm{Var}\ x_1)) \\
      &\mbf{in}\ x_2
    \end{align*}
    \caption{\label{subfig:tape-program-krawiec}
      Using \cite{ad-2021-krawiec-kmett-ad}
    }
  \end{subfigure}
  \end{center}
  \caption{\label{fig:tape-program}
    On the left, a small program producing the sharing structure of $\Contrib$ values for the program given in the text for our algorithm using defunctionalisation (\cref{sec:improve-defunctionalisation}).
    On the right, the output of differentiation of the same source program using \cite{ad-2021-krawiec-kmett-ad}.
  }
\end{figure}

After the defunctionalisation of \cref{sec:improve-defunctionalisation}, backpropagators have become $\Contrib$ values; however, the sharing structure is (of course) the same.
This new representation is given in \cref{subfig:sharing-structure-after}.
Assuming that $\mit{cb}_x$ and $\mit{cb}_y$ are the $\Contrib$ values for the inputs $x$ and $y$, respectively, the sharing graph in \cref{subfig:sharing-structure-after} could also be produced by the program in \cref{subfig:tape-program-dualrev}.
In a sense, this program is the fully flattened sequence of operations performed by the differentiated program (after defunctionalisation), ignoring all non-$\Contrib$ computation.
For $x$, $y$ and $z$ one should read the actual runtime values (of type $\R$) that were computed in the original program.

This trace is of a very standard shape; it records for each primitive operation executed:
\begin{itemize}
\item
  Which primitive operations produced the inputs to the operation (this is encoded in both the IDs (1, 2 and 1, 3) and the references to the previous $\Contrib$ values);
\item
  What the partial derivatives of the result of this operation are with respect to each of the inputs (these are the (1.0, 1.0 and $z$, $x$) values).
\end{itemize}
This is fairly standard for taping-based approaches to AD, because indeed, as we saw in the development in this paper, one can iterate in reverse over this array and propagate cotangents all the way to the input.

It is also the same shape as the algorithm described in~\cite{ad-2021-krawiec-kmett-ad}\footnote{This is a formalisation of the algorithm that Edward Kmett's \texttt{ad} Haskell library uses: \url{https://hackage.haskell.org/package/ad}} produces.
Indeed, differentiating the same original program using the algorithm in~\cite{ad-2021-krawiec-kmett-ad} yields the program in \cref{subfig:tape-program-krawiec}, where $\mrm{Add}$, $\mrm{Var}$, $\mrm{Scale}$ are constructors of a Delta data type that they define.
This representation is easily seen to encode the same information as \cref{subfig:tape-program-dualrev}.

The implications of this striking connection are left as future work.

\section{Extending the source language}\label{sec:source-language-extension}

The source language (\cref{fig:source-language}) that the algorithm discussed so far works on, is a higher-order functional language including product types and primitive operations on scalars.
However, dual-numbers reverse AD generalises to much richer languages in a very natural way, because most of the interesting work happens in the scalar primitive operations.

\paragraph{Recursion}
For example, we can allow recursive functions in our source language by adding the syntax $\mbf{letrec}\ f : \sigma \ra \tau = \fun(x : \sigma).\ s\ \mbf{in}\ t$.
The code transformation $\trans i{}$ for all $i$ then treats $\mbf{letrec}$ exactly the same as $\mbf{let}$---note that replacing $\mbf{letrec}$ by $\mbf{let}$, and vice-versa, produces valid syntax except for scoping---and the algorithm remains both correct and efficient.
In this way, the source language becomes Turing-complete.

\paragraph{Coproducts}
We can also easily extend the type system of the source language, for example by adding coproducts (sum types).
First add coproducts to the syntax for types ($\sigma,\tau \coloneqqq \cdots \mid \sigma + \tau$) both in the source language and in the target language, and add constructors and eliminators to all term languages (both linear and non-linear):
\begin{align*}
s,t \coloneqqq \cdots \mid \inl(t) \mid \inr(t) \mid \case s\ \{ \inl(x) \ra t_1 ; \inr(y) \ra t_2 \}
\end{align*}
where $x$ and $y$ are in scope in $t_1$ and $t_2$.
Then the type and code transformations extend in the unique structure-preserving manner:
\begin{gather*}
  \trans1c[\sigma + \tau] = \trans1c[\sigma] + \trans1c[\tau] \\
  \trans1c[\inl(t)] = \inl(\trans1c[t]) \qquad
    \trans1c[\inr(t)] = \inr(\trans1c[t]) \\
  \trans1c[\case s\ \{ \inl(x) \ra t_1 ; \inr(x) \ra t_2 \}] =
    \case{\trans1c[s]}\ \{ \inl(x) \ra \trans1c[t_1] ; \inr(x) \ra \trans1c[t_2] \}
\end{gather*}
The type transformation stays unchanged when moving to $\trans5{}$, and the only change for the term definitions is to transition to monadic code in $\trans3c$.
Lifting a computation to monadic code is a well-understood process.

Because we added a new type form to our language, we need to add an additional case to $\Interleave5$ and $\Deinterleave5$ as well:
\begin{align*}
  &\begin{array}{l@{\ }c@{\ }l}
    \Interleave5_\tau &:& \tau \ra \Int \ra ((\trans5{}[\tau], \IArray \R \ra \tau), \Int) \\
    \Interleave5_{\sigma + \tau} &=&
      \fun x.\ \fun i.\ \case x\ \{ \inl(y) \hspace{0.04cm} \ra \mbf{let}\ ((y', f), i') = \Interleave5_\sigma\ y\ i \\
      && \hspace{4.05cm} \mbf{in}\ ((\inl(y'), \fun \mit{arr}.\ \inl(f\ \mit{arr})), i') \\
      && \hspace{2.468cm} ; \inr(y) \ra \mbf{let}\ ((y', f), i') = \Interleave5_\tau\ y\ i \\
      && \hspace{4.05cm} \mbf{in}\ ((\inr(y'), \fun \mit{arr}.\ \inr(f\ \mit{arr})), i') \}
  \end{array} \\
  &\begin{array}{l@{\ }c@{\ }l}
    \Deinterleave5_\tau &:& \trans5{}[\tau] \ra (\tau, \tau \lra (\State \rlra \State)) \\
    \Deinterleave5_{\sigma + \tau} &=&
      \fun x.\ \case x\ \{ \inl(y) \hspace{0.05cm} \ra \mbf{let}\ (y', f) = \Deinterleave5_\sigma\ y \\
      && \hspace{3.590cm} \mbf{in}\ (\inl(y'), \fun d.\ \inl(f\ d)) \\
      && \hspace{2.005cm} ; \inr(y) \ra \mbf{let}\ (y', f) = \Deinterleave5_\tau\ y \\
      && \hspace{3.585cm} \mbf{in}\ (\inr(y'), \fun d.\ \inr(f\ d)) \}
  \end{array}
\end{align*}
As can be seen, all that happens is to write the only reasonable thing that is type-correct.

\paragraph{Recursive types}
In Haskell one can define (mutually) recursive data types e.g.\ as follows:
\begin{align*}
  &\mbf{data}\ T_1\ \alpha = C_1\ \alpha\ (T_2\ \alpha) \mid C_2\ \Int \\
  &\mbf{data}\ T_2\ \alpha = C_3\ \Int\ (T_1\ \alpha)\ (T_2\ \alpha)
\end{align*}
If the user has defined some data types, then on the condition that these data types do not contain explicit scalar values---i.e.\ they only contain scalars if some of their type parameters are instantiated with a type containing $\R$---we can allow these data types in the code transformation.
The \emph{declarations} of the data types (i.e.\ those shown above) are kept completely as-is; to the type transformation of the program itself, we add one rule for each data type that simply maps the type transformation over all type parameters (here just one, $\alpha$):
\begin{gather*}
  \trans1c[T_1\ \alpha] = T_1\ (\trans1c[\alpha]) \qquad
  \trans1c[T_2\ \alpha] = T_2\ (\trans1c[\alpha])
\end{gather*}
For example, if the codomain type of the program to differentiate was $\mathsf{Either}\ \R\ \Int$, we get $\trans1c[\mathsf{Either}\ \R\ \Int] = \mathsf{Either}\ (\R, \R \lra c)\ \Int$.
The restriction that data types used in the program to differentiate cannot mention $\R$ ensures that all scalars in the involved data types get fully differentiated by applying the type transformation at the top level only.\footnote{This restriction could be removed using a more involved transformation that produces declarations for new, differentiated data types for use in the differentiated program.}

The code transformation on terms is completely analogous to a combination of coproducts (given above in this section) and products (given already in \cref{fig:algo-naive}).
The wrapper also changes analogously: $\Interleave5$ and $\Deinterleave5$ obtain an additional clause for each monomorphic instantiation of the data types.
That is to say, we get clauses for $\Interleave5_{(T_1\,\R)}$, $\Interleave5_{(T_1\,(\Int, \R))}$, etc.
(This is a finite number of clauses because the program is not infinitely large, so does not mention an infinite number of distinct monomorphic types.\footnote{To support polymorphic recursion (e.g.\ as used in finger trees~\cite{fp-2006-finger-trees}), considering all monotypes does not suffice anymore; then, one can make each such clause take the (de)interleaving functions of all type parameters of the data type as arguments. Our implementation does not do this and hence does not (yet) support polymorphically recursive data types.})
In $\Wrap5$, where monomorphic types for $\sigma$ and $\tau$ are known, the correct version of $\Interleave5$ and $\Deinterleave5$ can then be invoked.

\paragraph{Conclusion}
In a sense, the efficiency of the algorithm is independent of the language constructs in the source language.
Indeed, in the forward pass, the code transformation is fully structure-preserving outside of the scalar constant and primitive operation cases; and in the reverse pass (in $\ResolveState$), all program structure is forgotten anyway, because the computation is flattened to a linear sequence of primitive operations on scalars.

\section{Implementation}\label{sec:implementation}

The advantage of a dual-numbers-based AD algorithm is that it scales well to a variety of language features.
In particular, it should apply without too much effort to a higher-order functional language with data types such as Haskell.
We have a prototype implementation\footnote{\url{https://github.com/tomsmeding/ad-dualrev-th}} of the algorithm in \cref{fig:algo-mutarrays}, together with the improvements from \cref{sec:improve-one-array,sec:improve-defunctionalisation}, that differentiates a basic fragment of Haskell (reinterpreted as a strict, call-by-value language) using Template Haskell.
The implementation does not incorporate the changes given in \cref{sec:improve-taping} that transform the algorithm into classical taping, but it does include support for recursive functions and user-defined data types as described in \cref{sec:source-language-extension}.

Template Haskell~\cite{fp-2002-template-haskell} is a built-in metaprogramming facility in GHC Haskell that (roughly) allows the programmer to write a Haskell function that takes a block of user-written Haskell code, do whatever it wants with the AST of that code, and finally splice the result back into the user's program.
The resulting code is still type-checked as usual.
The AST transformation that we implement is, of course, differentiation.

The full program transformation is defined in \texttt{src/Language/Haskell/ReverseAD/TH.hs}; property tests (comparing gradients against finite differencing) can be found in \texttt{test/Main.hs}.

\paragraph{Benchmarks}
To check that our implementation has reasonable performance in practice, we benchmark (in \texttt{bench/Main.hs}) against Kmett's \texttt{ad} library~\cite{ad-2021-kmett-hackage} (version 4.5) on a few basic functions.
These functions are:
\newcommand\rotatevecbyquat{\texttt{rotate\textunderscore{}vec\textunderscore{}by\textunderscore{}quat}}
\begin{itemize}
\item A single scalar multiplication of type \texttt{(Double, Double) -> Double};
\item Dot product of type \texttt{([Double], [Double]) -> Double};
\item Matrix--vector multiplication, then sum the result: of type \texttt{([[Double]], [Double]) -> Double};
\item The \rotatevecbyquat{} example from \cite{ad-2021-krawiec-kmett-ad} of type \texttt{(Vec3 Double, Quaternion Double) -> Vec3 Double}, with \texttt{data Vec3 s = Vec3 s s s} and \texttt{data Quaternion s = Quaternion s s s s}.
\end{itemize}
The last case has a non-trivial return type.

\begin{table}
\begin{tabular}{l|ccc}
  & TH & \texttt{ad} & TH / \texttt{ad} \\ \hline
  scalar mult. & 0.146 $\mu$s $\pm0.000$ & 0.536 $\mu$s $\pm0.002$ & \hphantom{$\approx$}0.27 \\
  dot product & 2.21\hphantom0 $\mu$s $\pm0.10$\hphantom0 & 2.07\hphantom0 $\mu$s $\pm0.06$\hphantom0 & $\approx$1.1\hphantom0 \\
  sum-mat-vec & 2.05\hphantom0 $\mu$s $\pm0.14$\hphantom0 & 1.32\hphantom0 $\mu$s $\pm0.05$\hphantom0 & $\approx$1.5\hphantom0 \\
  rotate\textunderscore{}vec\textunderscore{}by\textunderscore{}quat & 8.77\hphantom0 $\mu$s $\pm0.01$\hphantom0 & 6.13\hphantom0 $\mu$s $\pm0.02$\hphantom0 & $\approx$1.43 \\
\end{tabular}
\caption{\label{fig:bench-results}
  Benchmark results of \cref{fig:algo-mutarrays} + \cref{sec:improve-one-array,sec:improve-defunctionalisation} versus \texttt{ad-4.5}.
  The `TH' and `\texttt{ad}' columns indicate runtimes on one machine for our implementation and the \texttt{ad} library, respectively.
  The last column shows the ratio between the previous two columns.
  We give the size of the largest side of \texttt{criterion}'s 95\% confidence interval.
  Setup: GHC 9.2.2 on Linux, Intel i9-10900K CPU. Benchmarks are single-threaded.
}
\end{table}

The benchmark results are shown in \cref{fig:bench-results}.
The benchmarks are timed using the \texttt{criterion}\footnote{By Bryan O'Sullivan: \url{https://hackage.haskell.org/package/criterion}} library.
To get statistically significant results, measurements are performed by repeating benchmarks many times.
``Repeating a benchmark $n$ times'' means the following for our benchmarks:
\begin{itemize}
\item Scalar multiplication and \rotatevecbyquat{} are simply differentiated $n$ times;
\item Dot product is performed on lists of length $n$;
\item Matrix multiplication is performed on a matrix and vector of size $\sqrt n$, in order to get linear scaling in $n$.
\end{itemize}
By the results in \cref{fig:bench-results}, we see that for less trivial programs, our implementation is around 50\% slower than the highly-optimised \texttt{ad} library.
Because our code is simply a proof-of-concept, we conclude from this that the algorithm described in this paper indeed admits a work-efficient implementation.

\section{Conclusions \& future work}\label{sec:future-work}

We have presented a sequence of natural optimisation steps that transform a reverse AD algorithm that can be very rigorously proven correct but is very inefficient, to one that has the correct time complexity with a good constant factor.
However, there are a number of directions for improvement and further research.
\begin{itemize}
\item
  The sequentially generated integer IDs form a very crude over-approximation of the dependency structure of the backpropagators, namely to a linear graph.
  If the source program used parallel combinators such as \texttt{map}, \texttt{scan}, \texttt{fold}, etc.\ then the (statically-known) independence relations between computations performed by the many threads get lost in this linear approximation of the dependency graph.
  Hence, to preserve parallelism after differentiation, we need a more accurate representation of these dependency relationships.
  This means that the ordering between the tags of backpropagators becomes less trivial than a simple less-than operation on integers.
  How to preserve parallelism, especially for non-trivial parallel combinators such as \texttt{scan}, is yet unknown.

\item
  The primitive operations in our source language ($\Op_n$) are first-order and return a single scalar.
  This can be generalised somewhat beyond what we discuss in this paper, but it is unclear how to make the algorithm work in a compositional manner for \emph{higher-order} primitive operations such as numerical differential equation solvers or root finding routines, which are used in practice for e.g.\ statistical modelling \cite{appli-2021-disease-modelling-stan-tutorial,appli-2019-disease-modelling-stan-hmc-vi}, where reverse AD is also extensively used~\cite{ad-2015-stan-math-implementation}.

\item
  The link to the work of \cite{ad-2021-krawiec-kmett-ad} that was hinted at in \cref{sec:relation-krawiec} is tantalising, but not further explored in this paper.

\item
  While correctness proofs of the initial algorithm (\cref{fig:algo-naive}) exist~\cite{ad-2020-sam-mathieu-matthijs,nunes-2022-dual-numbers}, and while we argue intuitively for the correctness of our optimisation steps, we do not have a formal correctness proof of the final algorithm.
  We do believe that proving preservation of semantics of the optimisation steps will be easier than proving correctness of the final algorithm directly.

\item
  Similarly, we argue that the final algorithm has the correct computational complexity, but we do not have a formal proof.
\end{itemize}

\section{Related work}\label{sec:related-work}
The literature about automatic differentiation spans many decades and academic subcommunities (scientific computing, machine learning and -- most recently -- programming languages).
Important early references are \cite{ad-1964-ad,linnainmaa1970representation,ad-1980-ad}.
Good surveys can be found in \cite{ad-2018-survey-automatic-differentiation,ad-2018-survey-ad-implementation}.
In the rest of this section, we will focus on the more recent literature that studies AD from a programming languages (PL) point of view.

\subsection{Theoretical foundations for our algorithm}
The first mention that we know of the dual-numbers reverse mode AD algorithm ($\mathcal R_{3\text{dual}}$ of our \cref{sec:rev-ad-type}) that we analyse in this paper 
is \cite[page 12]{ad-2008-reverse-functional-ad}, where it is quickly dismissed before a different technique is pursued.
The algorithm is first thoroughly studied by \cite{ad-2020-dualnum-revad-linear-factoring}
using operational semantics and \cite[Section 6]{ad-2020-sam-mathieu-matthijs} using denotational semantics.
\cite{ad-2020-dualnum-revad-linear-factoring} introduces the key idea that underlies the efficient implementation of our paper: the linear factoring rule, stating that a term $f\ x + f\ y$, with $f$ a linear function, may be reduced to $f\ (x + y)$.
We build on their use of this rule as a tool in a complexity proof to make it a suitable basis for a performant implementation.

\cite{ad-2021-dual-revad-linear-factoring-pcf} extends the work of \cite{ad-2020-dualnum-revad-linear-factoring} to apply to a language with term recursion.
Similarly, \cite{nunes-2022-dual-numbers} extends the work of \cite{ad-2020-sam-mathieu-matthijs} to apply to recursive types, thus giving a correctness proof for initial dual-numbers reverse AD transformation of \cref{fig:algo-naive} when applied to idealised Haskell98.

These dual-numbers reverse AD algorithms arose from an attempt to pursue a reverse mode equivalent of the classical
forward mode AD algorithms based on dual numbers, such as the implementation of \cite{ad-2019-fwd-ad-gradient-compiler-opts}
and theoretical studies \cite{ad-2020-sam-mathieu-matthijs,vakar-staton-huot-2021,nunes-2022-dual-numbers}.

\subsection{Closely related algorithms }
We discuss in \cref{sec:relation-krawiec} how our technique relates to the implementation 
\cite{ad-2021-kmett-hackage} and its theoretical analysis in \cite{ad-2021-krawiec-kmett-ad}.
Further, in \cref{sec:improve-taping}, we explain the precise connection between our approach and classical taping 
approaches such as that taken by PyTorch \cite{ad-2017-pytorch}.

\cite{ad-2020-rev-ad-semantics} formalises the use of a classical taping approach when applied to a functional language as a custom operational semantics.
Their approach first reduces a functional program to an execution trace, essentially a graph of primitive operations, by using a form of symbolic execution and, next, differentiates that graph.
By contrast, our approach directly differentiates a functional source program using a code transformation that produces code that is efficient under a standard operational semantics.
In essence, we notice that it is not necessary to reduce to an execution trace before differentiating: we can simply associate all scalars with backpropagators and transform them under primitive operations to achieve a very similar tracing behaviour. 
Rather than suggesting an implementation like ours that exploits the linear factoring rule, \cite{ad-2020-dualnum-revad-linear-factoring} proposes to follow symbolic execution techniques like those of \cite{ad-2020-rev-ad-semantics} in an implementation.

\subsection{Other PL literature about AD}
\paragraph{CHAD and category theory inspired AD}
As we discuss in \cref{sec:rev-ad-type}, there is some freedom in how we choose the type of a 
reverse differentiated program.
The approach taken by Elliott \cite{adfp-2018-categories-ad}, $\mathcal R_2$ in our \cref{sec:rev-ad-type}, assigns a more precise type to the reverse 
derivative than we do. 
In particular, it enforces with the type system that the (primal) function value can only depend on the primal function input, not on the input cotangent.
Further, it pairs vectors (and values of other composite types) with a single composite backpropagator, rather than decomposing to the point where each scalar is paired with a mini-backpropagator like in our dual-numbers approach.
The resulting algorithm is extended to source languages with function types in 
\cite{vakar-2021-higher-order-reverse-ad,vakar-2022-chad,vytiniotis2019differentiable} and to sum and (co)inductive types in 
\cite{nunes-2022-chad-expressive}.
Like our dual-numbers reverse AD approach, the algorithm arises as a canonical structure-preserving functor on the syntax of a programming language.
However, due to a different choice in target category (a Grothendieck construction of a linear $\lambda$-calculus for CHAD rather than the syntax of a plain $\lambda$-calculus for dual-numbers AD), the resulting algorithm looks very different.

\paragraph{Speeding up derivatives at higher types using closure conversion}
In CHAD-like approaches to reverse AD, the obvious derivative of higher-order functions is inefficient.
The cause is that the separation of the derivative with respect to function arguments and captured context variables can lead to recomputation.
One solution to this problem, pioneered by \cite{ad-2008-reverse-functional-ad} and later 
analysed by \cite{vytiniotis2019differentiable,alvarez-picallo-2021}
is to use closure conversion to remove any capturing of context variables before applying CHAD to higher order source code.
The dual-numbers reverse AD approach of this paper does not suffer from this weakness and does not require closure conversion to achieve an efficient application to higher order source programs.

\paragraph{Approaches utilising non-local control flow}
Another category of approaches to AD recently taken by the PL community are those that 
rely on forms of non-local control flow such as delimited continuations \cite{ad-2018-rev-delimited-continuations}
or effect handlers \cite{sigal2021automatic,de2021verifying}.
These techniques are different in the sense that they generate code that is not purely functional. 
This use of non-local control flow makes it possible to achieve an efficient implementation of reverse AD that looks strikingly simple compared to alternative approaches.
Where the CHAD approaches and our dual-numbers reverse AD approach both have to manually invert control flow at compile time by making use of continuations that get passed around, combined with smart staging of execution of those continuations in our case, this work can be deferred to run time by clever use of delimited control operators or effect handlers.

\paragraph{Parallelism preserving AD systems}
Recently, 
\cite{dex-2021-ad,ad-2022-futhark-partial-recompute} have applied AD to functional source languages with parallel array operations.
A focus in their work has been to let the differentiated program inherit as much parallelism from the source program as possible. 
By contrast, in this work, we have exclusively focussed on differentiating sequentially executed code and, in fact, our approach using linearly ordered IDs is inherently sequential.
As we explain in \cref{sec:future-work}, we plan to investigate in future work whether the approach described in this paper can be made suitable for differentiating parallel programs.

\bibliography{bibliography}

\newpage
\appendix

\section{Type systems for source and target language}\label{app:type-system}

\newcommand\iru[3]{\frac{#1}{#2}}
\newcommand\Type{\mathsf{Type}}
\newcommand\TType{\mathsf{TType}}
\newcommand\PlainType{\mathsf{PDType}}

In this section we give an explicit type system for the source language in \cref{fig:source-language} and the initial target language in \cref{fig:target-language-1}.
This is mostly to show that, indeed, the typing is standard and unsurprising.

Types are determined by the $\Type$ judgement, and the subset of non-function types is determined by $\PlainType$ (``plain-data type'').
\begin{gather*}
\iru{}{\R\ \Type}{TReals} \qquad
\iru{}{()\ \Type}{TUnit} \qquad
\iru{}{\Int\ \Type}{TInt} \qquad
\iru{\sigma\ \Type \quad \tau\ \Type}{(\sigma, \tau)\ \Type}{TPair} \qquad
\iru{\sigma\ \Type \quad \tau\ \Type}{(\sigma \ra \tau)\ \Type}{TFun} \\
\iru{}{\R\ \PlainType}{TPReals} \qquad
\iru{}{()\ \PlainType}{TPUnit} \qquad
\iru{}{\Int\ \PlainType}{TPInt} \qquad
\iru{\sigma\ \PlainType \quad \tau\ \PlainType}{(\sigma, \tau)\ \PlainType}{TPPair}
\end{gather*}
Note that $\PlainType$ is indeed a subset of $\Type$.

For the typing of the lambda calculi, we define environments:
\begin{gather*}
\Gamma \coloneqq \varepsilon \mid \Gamma, x : \tau
\end{gather*}
and a judgement to check whether a binding is present in an environment:
\begin{gather*}
\iru{}{x : \tau \in \Gamma, x : \tau}{} \qquad
\iru{x : \tau \in \Gamma}{x : \tau \in \Gamma, y : \sigma}{}
\end{gather*}
A source language term $t$ has type $\tau$ in the environment $\Gamma$ if the judgement $\Gamma \vdash t \isS \tau$ holds.
\begin{gather*}
\iru{}{\Gamma \vdash r \isS \R}{} \qquad
\iru{x : \tau \in \Gamma}{\Gamma \vdash (x : \tau) \isS \tau}{} \qquad
\iru{}{\Gamma \vdash () \isS ()}{} \qquad
\iru{\Gamma \vdash s \isS \sigma \quad \Gamma \vdash t \isS \tau}{\Gamma \vdash (s, t) \isS (\sigma, \tau)}{} \\
\iru{\Gamma \vdash t \isS (\sigma, \tau)}{\Gamma \vdash \fst(t) \isS \sigma}{} \qquad
\iru{\Gamma \vdash t \isS (\sigma, \tau)}{\Gamma \vdash \snd(t) \isS \tau}{} \qquad
\iru{\Gamma \vdash s \isS \sigma \ra \tau \quad \Gamma \vdash t \isS \sigma}{\Gamma \vdash s(t) \isS \tau}{} \qquad
\iru{\sigma\ \Type \quad \Gamma, x : \sigma \vdash t \isS \tau}{\Gamma \vdash (\fun(x : \sigma).\ t) \isS \tau}{} \\
\iru{\sigma\ \Type \quad \Gamma \vdash s \isS \sigma \quad \Gamma, x : \sigma \vdash t \isS \tau}{\Gamma \vdash (\mbf{let}\ x : \sigma = s\ \mbf{in}\ t) \isS \tau}{} \qquad
\iru{\mit{op} \in \Op_n \quad \Gamma \vdash t_1 \isS \R \  \ldots \  \Gamma \vdash t_n \isS \R}{\Gamma \vdash \mit{op}(t_1,\ldots,t_n) \isS \R}{}
\end{gather*}
In the target language, types (judged by $\TType$) furthermore include linear functions between plain-data types:
\begin{gather*}
\iru{\tau\ \Type}{\tau\ \TType}{} \qquad
\iru{\sigma\ \PlainType \quad \tau\ \PlainType}{(\sigma \lra \tau)\ \TType}{}
\end{gather*}
Where we used $\isS$ for the source language, we use $\isT$ for the terms of the target language.
For linear function bodies, a term $b$ has type $\tau$ in environment $\Gamma$ with linear variable $z : \rho$ in scope if $\Gamma \vdash_{z : \rho} b \isB \tau$.
\begin{gather*}
\iru{\sigma\ \PlainType \quad \Gamma \vdash_{z : \sigma} b \isB \tau}{\Gamma \vdash (\lfun(z : \sigma).\ b) \isT \sigma \lra \tau}{} \\
\text{(Plus all the $\isS$ rules with $\isS$ replaced by $\isT$.)} \\
\iru{}{\Gamma \vdash_{z:\tau} z \isB \tau}{} \qquad
\iru{\Gamma \vdash_{z:\rho} b \isB \tau \quad \Gamma \vdash_{z:\rho} b' \isB \tau}{\Gamma \vdash_{z:\rho} b + b' \isB \tau}{} \qquad
\iru{}{\Gamma \vdash_{z:\rho} \underline0 \isB \tau}{} \\
\iru{}{\Gamma \vdash_{z:\rho} () \isB ()}{} \qquad
\iru{\Gamma \vdash_{z:\rho} s \isB \sigma \quad \Gamma \vdash_{z:\rho} t \isB \tau}{\Gamma \vdash_{z:\rho} (s, t) \isB (\sigma, \tau)}{} \qquad
\iru{\Gamma \vdash_{z:\rho} t \isB (\sigma, \tau)}{\Gamma \vdash_{z:\rho} \fst(t) \isB \sigma}{} \qquad
\iru{\Gamma \vdash_{z:\rho} t \isB (\sigma, \tau)}{\Gamma \vdash_{z:\rho} \snd(t) \isB \tau}{} \\
\iru{x : \sigma \lra \tau \in \Gamma \quad \Gamma \vdash_{z:\rho} b \isB \sigma}{\Gamma \vdash_{z:\rho} (x : \sigma \lra \tau)(b) \isB \tau}{} \qquad
\iru{\mit{op} \in \Op_n \quad 1 \leq i \leq n \quad x_1 : \R \in \Gamma \  \ldots \  x_n : \R \in \Gamma \quad \Gamma \vdash_{z:\rho} b \isB \R}{\Gamma \vdash_{z:\rho}\partial_i\mit{op}(x_1,\ldots,x_n)(b) \isB \R}{}
\end{gather*}

\end{document}